\documentclass[12]{iopart}

  \usepackage{amsthm}
  \usepackage{iopams}
  \usepackage{setstack}
  \usepackage{bbm}
  \usepackage{fancyhdr}
  \usepackage{enumitem}
  \usepackage{graphicx}
  \usepackage[subrefformat=parens]{subfig}
  \usepackage{nicefrac}

  \newcommand{\NN}{\mathbb N}   
  \newcommand{\ZZ}{\mathbb Z}   
  \newcommand{\CC}{\mathbb C}   
  \newcommand{\EE}{\mathbb E}   
  \newcommand{\cC}{\mathcal C}  
  \newcommand{\cD}{\mathcal D}  
  \newcommand{\cE}{\mathcal E}  
  \newcommand{\cL}{\mathcal L}  
  \newcommand{\cP}{\mathcal P}  
  \newcommand{\cS}{\mathcal S}  
  \newcommand{\cV}{\mathcal V}  
  \newcommand{\id}{\mathbbm 1}  
  \newcommand{\gL}{\mathfrak L} 
  
  \newcommand{\lf}{\left\lfloor} 
  \newcommand{\rf}{\right\rfloor}
  
  \newcommand{\lv}{\left|} 
  \newcommand{\rv}{\right|}

  \newcommand{\ds}{\displaystyle}
  
  \newcommand{\Mod}[1]{\;(\bmod\; #1)} 
  \newcommand{\dif}{\,\textup{d}} 
  \newcommand{\derive}[2][]%
      {\frac{\textup{d} #1}{\textup{d} #2}} 
  \newcommand{\pderive}[2][]%
      {\frac{\partial #1}{\partial #2}} 
  \newcommand{\nderive}[3][]%
      {\frac{\textup{d}^{#3} #1}{\textup{d} #2^{#3}}} 
  \newcommand{\npderive}[3][]%
      {\frac{\partial^{#3} #1}{\partial #2^{#3}}} 
  \renewcommand\vec{\bi}

  \newcommand{\csch}{\textrm{csch}}

\theoremstyle{plain}
  \newtheorem{thm}{Theorem}
  \newtheorem{lem}[thm]{Lemma}
  
  \newtheorem{cor}[thm]{Corollary}
\theoremstyle{definition}

\theoremstyle{remark}
  
\begin{document}

\title{Spectral properties of quantum circulant graphs}

\author{J M Harrison and E Swindle}

\address{Department of Mathematics, Baylor University, Waco, TX 76798, USA}
\ead{\mailto{jon\_harrison@baylor.edu}, \mailto{erica\_swindle@baylor.edu}}


\begin{abstract}
We introduce a new model for investigating spectral properties of quantum graphs, a quantum circulant graph. Circulant graphs are the Cayley graphs of cyclic groups. Quantum circulant graphs with standard vertex conditions maintain important features of the prototypical quantum star graph model. In particular, we show the spectrum is encoded in a secular equation with similar features. The secular equation of a quantum circulant graph takes two forms depending on whether the edge lengths respect the cyclic symmetry of the graph. When all the edge lengths are incommensurate, the spectral statistics correspond to those of random matrices from the Gaussian Orthogonal Ensemble according to the conjecture of Bohigas, Giannoni and Schmit.  When the edge lengths respect the cyclic symmetry the spectrum decomposes into subspectra whose corresponding eigenfunctions transform according to irreducible representations of the cyclic group.  We show that the subspectra exhibit intermediate spectral statistics and analyze the small and large parameter asymptotics of the two-point correlation function, applying techniques developed from star graphs. The particular form of the intermediate statistics differs from that seen for star graphs or Dirac rose graphs.  As a further application, we show how the secular equations can be used to obtain spectral zeta functions using a contour integral technique. Results for the spectral determinant and vacuum energy of circulant graphs are obtained from the zeta functions.
\end{abstract}

\vspace{2pc}

\submitto{\jpa}

\maketitle



\section{Introduction}

Quantum graphs were introduced as a model in quantum chaology by Kottos and Smilansky \cite{KS97,KS99}.  They have also been used to investigate Anderson localization, carbon nanotubes, graphene, photonic crystals, superconductivity and waveguides, see Berkolaiko and Kuchment \cite{BK13}, Gnutzmann and Smilansky \cite{GS06} or  Kuchment \cite{K04} for an introduction.

\begin{figure}[htb!]
\centering
  \subfloat[][]{
    \includegraphics{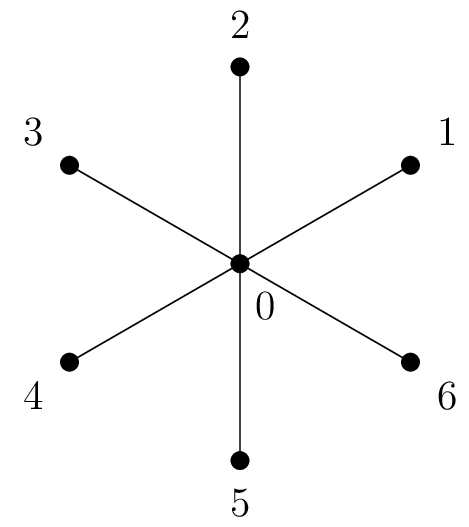}}
  \qquad\quad
  \subfloat[][]{
    \includegraphics{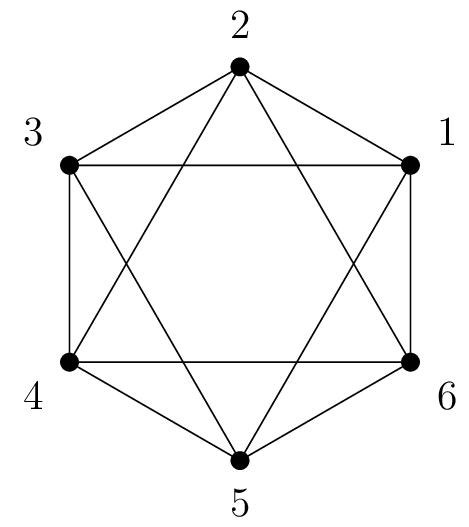}}
\caption[]{(a) A star graph with $6$ edges. (b) The circulant graph, $C_6(1,2)$.} \label{fig:quantumgraphs}
\end{figure}

Star graphs, graphs with a single central vertex connected to any number of surrounding vertices of degree one, see figure \ref{fig:quantumgraphs}(a), are a prototypical graph model. Every graph is locally a star graph at each vertex. Consequently, they are typically one of the first models to be investigated. When studying spectral properties, results are often obtained for star graphs by exploiting the simple structure. So results are known for star graphs that have yet to be proved for other classes of graphs and results are often obtained first for star graphs, see for example \cite{BK99,BKW04}.

One feature that makes the star graph particularly amenable to investigation is that its spectrum is encoded in a simple secular equation. The secular equation for the Laplace operator on a star graph with standard conditions at the vertices is,
\begin{equation}\label{starsecular}
\sum_{j=1}^E \tan k L_j = 0 \ ,
\end{equation} 
where the star has $E$ edges each of length $L_j$. When $k$ is a solution of the secular equation \eref{starsecular} then $k^2$ is an eigenvalue of the Laplacian.  

This article introduces a new quantum graph model that develops the quintessential star graph paradigm, quantum circulant graphs. A circulant graph, see figure \ref{fig:quantumgraphs}(b), is a Cayley graph of a cyclic group, see \sref{sec:circulant} for the definition.   We study two classes of quantum circulant graphs depending on whether the graph metric respects the cyclic symmetry of the graph.  The spectral results we obtain are based on two secular equations we develop for the two forms of the metric, theorems \ref{thm:seculareq} and \ref{thm:symseculareq}.  Importantly, these secular equations  maintain significant features of the star graph formula (\ref{starsecular}) which we 
exploit subsequently to obtain the form of the intermediate statistics, section \ref{sec:intermediate}, and to formulate spectral zeta functions, section \ref{sec:zeta} and following.

The distribution of the diameter of random circulant graphs was recently studied by Marklof and Str{\"o}mbergsson \cite{MS13}. While the distribution of diameters is a purely graph theoretic property, their approach, identifying circulant graphs with lattice graphs, has an analogy in the quotient graph technique used to identify intermediate statistics in systems where the metric respects the cyclic symmetry of the graph, see \sref{sec:quotientgraph}. The quotient graph approach was introduced by Band et al. in \cite{BPB09} and developed in \cite{PB10}.

Graph models have been used to investigate intermediate statistics in quantum systems which are classically chaotic \cite{BBK01,BK99,HW12}.  A classically chaotic quantum system with time-reversal symmetry has a spectrum whose statistical properties are conjectured to correspond to those of eigenvalues of large random matrices from the Gaussian Orthogonal Ensemble (GOE) \cite{BGS84}, a central tenet in the study of quantum chaology.  Intermediate spectral statistics differ from random matrix statistics although with some broad features in common, like linear level repulsion and the exponential decay of the probability of large spacings \cite{GMS01}.
For quantum circulant graphs the symmetry means that the spectrum decomposes into subspectra transforming according to irreducible representations of the cyclic group.
We show that, as in the case of star graphs, the subspectra associated with quantum circulant graphs with a symmetric metric exhibit intermediate spectral statistics.  However, (except for the subspectrum associated with the trivial representation) the exact form differs from that observed on star graphs or in the case of the Dirac operator on a rose graph both in the small and large parameter asymptotics of the pair correlation function.

The Ihara zeta function is an important tool in the spectral theory of combinatorial graphs \cite{I66,ST96,S86}.  For a quantum graph the spectrum $0 \leq \lambda_0 \leq \lambda_1 \leq \lambda_2 \leq \cdots$ can be encoded in a spectral zeta function which is formally,
\begin{equation}\label{spectralzetaintro}
\zeta(s) = \sideset{}{'} \sum_{i=0}^\infty \lambda_i^{-s},
\end{equation}
where the prime denotes that eigenvalues of zero are omitted.  The spectral zeta function can be obtained from the secular equations, theorems \ref{thm:seculareq} and \ref{thm:symseculareq}, following a contour integral technique introduced by Kirsten and McKane \cite{KM03,KM04} and applied to graphs by Harrison, Kirsten and Texier \cite{HK11,HKT12}. 
From the spectral zeta function we derive the vacuum energy and spectral determinants of the circulant graph models, both topics of research in their own right \cite{Aetal00,D00}.   
The vacuum energy of a graph was introduced as a model with a repulsive Casimir force on a star graph by Fulling and Wilson \cite{FKW07}. Further vacuum energy results for graphs are described in \cite{BHJ12,BHW09,HK12}.

The paper is arranged in the following way. In section \ref{sec:circulant} we introduce the two models of quantum circulant graphs that we investigate. In section \ref{sec:secular} we derive the secular equation of circulant graphs with incommensurate edge lengths. If instead the edge lengths are chosen to reflect the symmetry of the cyclic group, we show the secular equation factorizes into a product of factors, where each factor has the form of the sum in (\ref{starsecular}), section \ref{sec:symmetricsecular}.  
Hence the spectrum of circulant graphs with edge symmetry decomposes into subspectra whose eigenfunctions transform according to irreducible representations of the cyclic group.  In section \ref{sec:intermediate}, we apply techniques developed for star graphs to analyze the two-point correlation function and show that the subspectra exhibit a new form of intermediate statistics.  
The two forms of the secular equation allow the graph eigenvalues to be calculated efficiently and numerical examples are presented in \sref{sec:numerics}. Using the secular equations we employ a contour integral technique to derive the related spectral zeta functions, section \ref{sec:zeta}. Results for the spectral determinant obtained from the zeta function are presented in section \ref{sec:determinant} and section \ref{sec:vacuum} has equivalent results for the vacuum energy. We discuss extensions of the results presented in section \ref{sec:conclusions}.

\section{Circulant graphs}\label{sec:circulant}

There are several equivalent characterizations of circulant graphs, see e.g. \cite{Vilfred04}. For the purpose of this paper, we use a constructive version. Let $\cV = \left\{1,\dots,n \right\}$ be a set of $n$ vertices and $\cE$ denote a set of edges. An edge $e=(i,j)\in \cE$ is a pair of vertices. We write $i\sim j$ if the vertex $i$ is connected to $j$ by an edge. To construct a circulant graph, $C_n(\vec{a}) = (\cV,\cE)$, fix a vector of integers $\vec{a} = \left(a_1,\dots,a_d \right)$, such that $0 < a_1 < a_2 < \dots < a_d < n/2$.
  In general, it is possible that $a_d = n/2$, resulting in a double-edge, but we omit this case for simplicity.
Two vertices $i,j\in \cV$ are connected by the edge $(i,j)\in \cE$ if and only if $|i-j|\equiv a_h \Mod n$ for some $a_h\in \vec{a} $. Figure \ref{fig:quantumgraphs}(b) shows the circulant graph $C_6(1,2)$.

When $a_d<n/2$, circulant graphs are $2d$-regular, since each $a_h\in \vec{a}$ contributes $n$ edges. Additionally, $C_n(\vec{a})$ is connected if and only if $\gcd(a_1,\dots,a_d,n) = 1$. We assume this holds for all circulant graphs that we study.

The adjacency matrix of a circulant graph is a circulant matrix.
  An $m\times m$ \emph{circulant matrix} $C$ is uniquely determined by a row vector $\vec{c} = \left(c_1,\dots,c_m \right)$. The vector $\vec{c}$ becomes the first row of $C$ and every subsequent row of $C$ is a cyclic permutation of $\vec{c}$, with each element $\vec{c}$ moved one place to the right relative to the previous row. The \emph{representer} of $C$ is the polynomial,
  \begin{equation}\label{representer}
  p(z) = c_1 + c_2 z + \cdots + c_m z^{m-1} \ .
  \end{equation}

The determinant of a circulant matrix can be written in terms of the representer, see e.g. \cite{Davis79}.
\begin{equation}\label{circdet}
\det C = \prod_{j=1}^m p \left(\omega^{j-1} \right) \ ,
\end{equation}
where $\omega = \exp \left(2\pi\rmi / m \right)$. In the case of real-valued matrix elements this simplifies to,
\begin{equation}\label{realcircdet}
\det C = \cases{
         p(1) \prod_{j=1}^{(m-1)/2} \lv p \left(\omega^j \right)\rv^2 & for odd $m$ \\
         p(1) p(-1) \prod_{j=1}^{(m/2)-1} \lv p \left(\omega^j \right)\rv^2 & for even $m$
         }\ .
\end{equation}

\subsection{Quantum circulant graphs}

In a quantum graph, each edge corresponds to an interval and we consider a differential operator acting on functions on the set of intervals. Let $[0,L_{i,j}]$ be the interval associated with the edge $(i,j)\in \cE$, such that $i$ corresponds to $0$ and $j$ to $L_{i,j}$, or vice versa. The results are independent of the choice of orientation of the intervals. 
$L_{i,j}$ is referred to as the \emph{length} of the edge $(i,j)$ and we use the notation $C_n(\vec{L};\vec{a})$ for a circulant graph endowed with this metric.  The \emph{total length} of the graph is $\cL=\sum_{(i,j)\in \cE} L_{i,j}$. 

Functions $\psi$ on $C_n(\vec{L};\vec{a})$ are $(nd)$-tuples of functions $\psi_{i,j}\in L^2[0,L_{i,j}]$. The Hilbert space of the graph is then,
\begin{equation}\label{hilbertspace}
L^2 \left(C_n(\vec{L};\vec{a})\right) := \bigoplus_{(i,j)\in \cE} L^2 \Big([0,L_{i,j}]\Big) \ .
\end{equation}
  
In this article, we study the spectrum of the Laplace operator $\Delta$ on $C_n(\vec{L};\vec{a}) $. On each edge an eigenfunction $\psi_{i,j}$ with eigenvalue $k^2$ satisfies the equation, 
\begin{equation}\label{schrodingeredge}
-\nderive{x_{i,j}}{2} \psi_{i,j} = k^2 \psi_{i,j} \ .
\end{equation}
The domain of the operator is the set of functions in the second Sobolev space $H^2[0,L_{i,j}]$ on each edge, which satisfy the standard (or Neumann-like) vertex conditions,
\begin{eqnarray}
\psi_{i,j_1}(i) = \psi_{i,j_2}(i), \quad \textrm{ for all } j_1, j_2 \sim i \ , \label{vertexcondition1} \\
\sum_{j\sim i} \psi'_{i,j}(i) = 0 \ . \label{vertexcondition2}
\end{eqnarray}
We use $\psi_{i,j}(i)$ to denote the value of $\psi_{i,j}$ at the vertex $i$ which corresponds either to $x_{i,j}=0$ or $x_{i,j}=L_{i,j}$. The derivative in \eref{vertexcondition2} is taken in the outgoing direction from the vertex, so it is either $\psi'_{i,j}(0)$ or $-\psi'_{i,j}(L_{i,j})$. The negative Laplacian is self-adjoint on the given domain \cite{BK13}.

An alternative set of vertex conditions for which the Laplace operator on the metric graph is self-adjoint are Dirichlet conditions,
\begin{equation}\label{vertexconditionDirichlet}
\psi_{i,j}(i) = \psi_{i,j}(j) = 0 , \quad \textrm{ for all } (i,j)\in \cE \ .
\end{equation}
A function satisfying Laplace's equation on $[0,L_{i,j}]$ with Dirichlet boundary conditions is $\psi_{i,j} (x_{i,j}) = \sin(m\pi x_{i,j} / L_{i,j}) $. So with Dirichlet conditions the graph is broken up into a collection of independent intervals and the spectrum is the union of the spectra on the intervals.  We will refer to this as the \emph{Dirichlet spectrum} of the graph,
\begin{equation}\label{Dirichleteigenvalues}
\cD = \left\{\frac{m\pi}{L_{i,j}}: m\in \NN, (i,j)\in \cE \right\} \ .
\end{equation}

We will also consider a more specialized metric for $C_n(\vec{L};\vec{a})$ which respects the symmetry of $C_n(\vec{a})$. Fix a vector $\bell = \{\ell_1,\dots,\ell_h \}$ and assign edge lengths to $C_n(\vec{L};\vec{a})$ such that the edge $(i,j)$ has length $\ell_h$ whenever $|i-j|\equiv a_h \Mod n$. We let $C_n(\bell;\vec{a})$ denote a circulant graph with symmetric edge lengths.

\section{Secular equation}\label{sec:secular}

Within the study of spectral properties of quantum graphs, an important tool is a secular equation.  A secular equation is an equation whose roots correspond to the graph eigenvalues. In general, one may write a secular equation for a quantum graph using the $2\lv\cE\rv \times 2\lv\cE\rv$ bond scattering matrix \cite{BK13}.
In the case of circulant graphs, however, we can adapt techniques used to study star graphs \cite{BBK01} to develop a more explicit secular equation.

Let $\bPhi = \left(\phi_1,\dots,\phi_n \right)^T$ be a vector of the values of $\psi$ at each of the vertices, $\phi_j = \psi_{i,j}(j)$. Then the solution to \eref{schrodingeredge} on the edge $(i,j)$ can be written
\begin{equation}\label{edgesolution}
\psi_{i,j}(x) = \left(\frac{\phi_j - \phi_i \cos k L_{i,j}}{\sin k L_{i,j}}\right) \sin k x + \phi_i \cos k x \ ,
\end{equation}
provided $ \sin k L_{i,j} \neq 0 $. The vector $\bPhi$ defines an eigenfunction on $C_n(\vec{L};\vec{a})$ if the functions $\psi_{i,j}$ in \eref{edgesolution} satisfy the vertex conditions. Substituting in \eref{vertexcondition2} at vertex $i$,
\begin{equation}\label{fixedvertex}
\sum_{j\sim i} \left(\phi_j \csc k L_{i,j} - \phi_i \cot k L_{i,j} \right) = 0 \ , 
\end{equation}
which holds independent of the orientation of the intervals. Hence we have a system of $n$ equations \eref{fixedvertex}, one for each vertex. Writing the system in matrix form
\begin{equation}\label{matrixeq}
M(k) \bPhi = \mathbf{0}
\end{equation}
where $ M(k) = [m_{ij}]$ is the $n \times n$ matrix
\begin{eqnarray}
m_{ii} &= -\sum_{j\sim i} \cot k L_{i,j} \label{Mdiag} \\
m_{ij} &= \cases{
           \csc k L_{i,j} & if $i \sim j$ \\
           0              & if $i \nsim j$ \\
           } \ . \label{Moffdiag}
\end{eqnarray}
$\bPhi$ is a non-trivial vector in the null space of $M(k)$ if and only if 
\begin{equation}\label{seculareq}
\det M(k) = 0 \ ,
\end{equation}
which is our secular equation.

\begin{thm}\label{thm:seculareq}
  Let $k\in \CC \setminus \cD$. Then $\lambda = k^2$ is an eigenvalue of the negative Laplace operator on $C_n(\vec{L};\vec{a})$ with multiplicity $m$ if and only if $k$ is an $m$'th root of,
  \begin{equation*}
  \det M(k) = 0 \ .
  \end{equation*}
\end{thm}

\section{Secular equation with symmetric edge lengths}\label{sec:symmetricsecular}

For a quantum graph $C_n(\bell;\vec{a})$ with symmetric edge lengths $M(k)$ is 
\begin{eqnarray}
m_{ii} = -2 \sum_{h=1}^d \cot k \ell_{h} \label{Mdiagsym} \\
m_{ij} = \cases{
           \csc k \ell_h & if $|i-j| \equiv a_h \Mod n$ \\
           0             & otherwise \\
           } \ .\label{Moffdiagsym}
\end{eqnarray}
\begin{lem}
  Let $C_n(\bell;\vec{a})$ be a quantum circulant graph with an edge-symmetric metric. Then $M(k)$ is a circulant matrix. 
\begin{proof}
  For each $a_h\in \vec{a}$, let $\Gamma_h := C_n(a_h)$ be the subgraph containing only edges connected via $a_h$, and let $J_h$ be its adjacency matrix. Each $\Gamma_h$ is also a circulant graph, so each $J_h$ is a circulant matrix.
  \begin{equation}\label{Mcirculant}
  M(k) = \left(-2 \sum_{h=1}^d \cot k \ell_h \right) \id_n + \sum_{h=1}^d \left(\csc k \ell_h \right) J_h 
  \end{equation}
  As a linear combination of circulant matrices is a circulant matrix,  $M(k)$ is a circulant matrix. 
\end{proof}
\end{lem}
A representer of a circulant matrix was defined in \eref{representer}.
\begin{cor}
  The representer for the circulant matrix $M(k)$ of a quantum circulant graph with an edge-symmetric metric is
\begin{equation}\label{Mrepresenter}
p(z) = \sum_{h=1}^d \left[-2\cot(k\ell_h) + \left(z^{a_h} + z^{n-a_h} \right) \csc(k\ell_h) \right] \ .
\end{equation}
\end{cor}

In order to apply these results to the secular equation, we introduce functions $p_j(k) $ by evaluating $p$ at $\omega^j$ where $\omega = \exp \left(2\pi\rmi / n \right)$.
\begin{equation}\label{pj}
\fl
p_j(k) := 
          \cases{
          2\sum_{h=1}^d \tan\left(\frac{k \ell_h}{2}\right) & $j=0$ \\
          2\sum_{h=1}^d \left[-\cot(k\ell_h) + \left(\cos\left(\frac{2 \pi j a_h}{n}\right)\right) \csc(k\ell_h)\right] & $1\leq j\leq \lf \frac{n-1}{2} \rf$ \\
          2\left[\sum_{\substack{h=1 \cr a_h\textrm{ is even}}}^d \tan\left(\frac{k\ell_h}{2}\right) - \sum_{\substack{h=1 \cr a_h \textrm{ is odd}}} \cot\left(\frac{k\ell_h}{2}\right)\right] & $j = n/2$ \\
          } 
\end{equation}
We now have the following result.

\begin{thm}\label{thm:symseculareq}
Let $C_n(\bell;\vec{a})$ be a quantum circulant graph with symmetric edge lengths.
  \begin{enumerate}[label=\Roman*.]
  \item Let $n$ be odd and $k\in \CC \setminus \cD$.  Then $\lambda = k^2>0$ is an eigenvalue of the  negative Laplace operator on $C_n(\bell;\vec{a})$ with multiplicity $m$ if and only if $k$ is an $m$'th root of,
    \begin{equation}
    p_0(k) \ds\prod_{j=1}^{(n-1)/2} \lv p_j(k) \rv^2 = 0 \ .
    \end{equation}
  \item Let $n$ be even and $k\in \CC \setminus \cD$.  Then $\lambda = k^2>0$ is an eigenvalue of the  negative Laplace operator on $C_n(\bell;\vec{a})$ with multiplicity $m$ if and only if $k$ is an $m$'th root of,
    \begin{equation}
    p_0(k) p_{n/2}(k) \ds\prod_{j=1}^{(n/2)-1} \lv p_j(k) \rv^2 = 0 \ .
    \end{equation}
  \end{enumerate}
\end{thm}
%


\subsection{Quotient graph}\label{sec:quotientgraph}

The secular equation described for the quantum circulant graph with symmetric edge lengths in theorem \ref{thm:symseculareq} was obtained from the secular equation of a general quantum circulant graph, theorem \ref{thm:seculareq}. Alternatively the spectrum with symmetric edge lengths can be obtained from the quotient graph, introduced by Band, Parzanchevski and Ben-Shach in \cite{BPB09,PB10}.

A symmetry of a quantum graph is an automorphism of the metric graph which preserves both the edge lengths and vertex conditions.  For the circulant graph such a symmetry is the rotation $\sigma$ where $\sigma(v) = v+1$. The group generated by $\sigma$ is the cyclic group $\ZZ_n$. Irreducible complex representations of the cyclic group are $\cS_j$, with $j = 0,\dots,n-1$, where,
\begin{equation}
\cS_j(\sigma) = \rme^{\rmi \theta_j} \ ,
\end{equation}
for $\theta_j=2\pi j / n$.  

Let the eigenspace of the graph with eigenvalue $\lambda$ be,
\begin{equation}
\Psi(\lambda) = \ker(-\Delta - \lambda\id) \ .
\end{equation}
Then $\Psi(\lambda)$ can be decomposed into subspaces each transforming according to an irreducible representation $\cS_j$,
\begin{equation}
\Psi(\lambda) = \sum_{j=0}^{n-1} \Psi_j(\lambda) \ ,
\end{equation}
where for an eigenfunction $\psi$ in $\Psi_j(\lambda)$, 
\begin{equation}
\psi(\sigma(x)) =  \rme^{\rmi \theta_j} \psi(x) \ .
\end{equation}

\begin{figure}[htb!]
\centering
  \subfloat[][]{
    \includegraphics{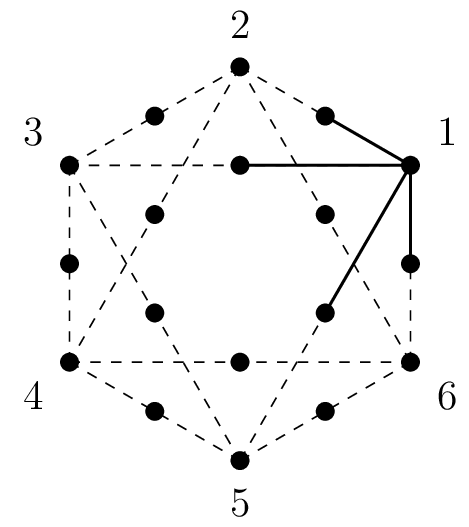}} 
  \qquad\quad
  \subfloat[][]{
    \includegraphics{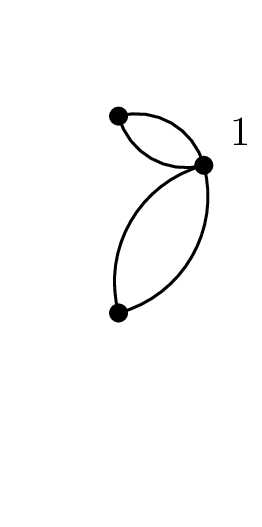}} 
\caption[]{(a) The circulant graph, $C_6(\bell;(1,2))$ with dummy vertices, the edges in the fundamental domain are shown with solid lines. (b) The corresponding quotient graph.}\label{fig:dummyvertices}
\end{figure}

For a symmetric circulant graph  $C_n(\bell;\vec{a})$ and representation $\cS_j$ one can construct a quotient graph $C_n(\bell;\vec{a}) / \cS_j$ following the procedure defined in \cite{BPB09,PB10}. Eigenfunctions of the quotient graph will be in bijection with the eigenfunctions on the circulant graph transforming according to a given irreducible representation $\cS_j$. To construct the quotient graph on each edge of the circulant graph we introduce a dummy vertex at $x = \ell_h / 2$ of degree two with Neumann-like conditions. A function on the graph is continuous with a continuous first derivative at the new vertices.  We will keep the original coordinates when introducing the dummy vertices so an edge corresponding to the interval $[0,\ell_h]$ is simply broken into subintervals $[0,\ell_h / 2]$ and $[\ell_h / 2,\ell_h]$ with a dummy vertex of degree two joining them. Introducing such dummy vertices does not change the spectrum of the original graph and eigenfunctions on the original graph are obtained from eigenfunctions on the subdivided graph by joining corresponding edge pairs. Figure \ref{fig:dummyvertices}(a) shows a circulant graph with dummy vertices.
 
The subgraph consisting of vertex $1$ with the neighboring dummy vertices forms a \emph{fundamental domain}. The whole circulant graph is generated by the action of $\sigma$ on the fundamental domain. In the fundamental domain there are two edges with length $\ell_h / 2$ for each $h$.  The quotient graph is constructed by identifying the two dummy vertices on edges of length $\ell_h/2$ in the fundamental domain,  see figure \ref{fig:dummyvertices}(b). The quotient graph then has $d$ vertices of degree $2$.  At these vertices we must introduce new vertex conditions which depend on the representation $\cS_j$.

For a function $\psi$ on the quotient graph, let $\psi_{h-}$ denote the function on the interval $[0,\ell_h / 2]$ and $\psi_{h+}$ the function on $[\ell_h / 2,\ell_h]$. Then at the $h$'th degree two vertex,
\begin{eqnarray}
\psi_{h-}(\ell_h/2) = \rme^{\rmi a_h \theta_j} \psi_{h+}(\ell_h/2) \ , \\
\psi'_{h-}(\ell_h/2) = \rme^{\rmi a_h \theta_j} \psi'_{h+}(\ell_h/2) \ .
\end{eqnarray}
The conditions at the central vertex of the quotient graph are the standard vertex conditions \eref{vertexcondition1} and \eref{vertexcondition2}, as in the original circulant graph. Then the spectrum of the quotient graph is the subset of the spectrum of the circulant graph whose eigenfunctions transform according to the irreducible representation $\cS_j$.  

%
%

A solution to \eref{schrodingeredge} on the pair of edges $h\pm$ can be written as a function defined piecewise on $[0,\ell_h]$,

\begin{equation}\label{piecewisesolngen}
\fl
\psi_h(x) = \cases{
            \alpha_h \cos(k x) + \beta_h \sin(k x) & if $0 \leq x < \ell_h / 2$ \\
            \alpha_h \rme^{-\rmi a_h \theta_j} \cos(k x) + \beta_h \rme^{-\rmi a_h \theta_j} \sin(k x) & if $\ell_h / 2 < x \leq \ell_h$ \\
            } \ .
\end{equation}
Let $\phi$ be the value of an eigenfunction of the quotient graph at the central vertex. Assuming $k \neq m\pi / \ell_h$, $m=1,2,\dots $
\begin{equation}\label{piecewisesoln}
\fl
\psi_h(x) = \left\{ \begin{array}{l}
            \phi \left[\cos(k x) + (\rme^{\rmi a_h \theta_j } \csc(k \ell_h ) - \cot(k \ell_h)) \sin(k x)\right] \\ 
            \qquad \qquad \qquad \qquad \qquad  \qquad \qquad \qquad \quad \textrm{if }0 \leq x < \ell_h / 2 \\
            \phi \rme^{-\rmi a_h \theta_j} \left[\cos(k x) + (\rme^{\rmi a_h \theta_j } \csc(k \ell_h ) - \cot (k \ell_h)) \sin(k x)\right] \\
            \qquad \qquad \qquad \qquad \qquad  \qquad \qquad \qquad \quad \textrm{if }\ell_h / 2 < x \leq \ell_h  \\
            \end{array}\right. \ .
\end{equation}
Condition \eref{vertexcondition2} at the central vertex requires,
\begin{equation}\label{quotientvertexcondition2}
\sum_{h=1}^d \psi_h'(0)-\psi_h'(\ell_h) =0 \ .
\end{equation}
Then for $\phi \neq 0$,
\begin{equation}\label{eq:quotientsecular} 
2\sum_{h=1}^d \cos(a_h \theta_j) \csc(k \ell_h) - \cot(k \ell_h) = 0 \ ,
\end{equation}
which is equivalently $p_j(k) = 0$ for $0 \leq j \leq \lf n / 2 \rf$. For $j > n / 2$, we define $p_j(k) := p_{n-j}(k)$ since $\theta_{n-j} = 2\pi - \theta_j$. This is a secular equation for eigenvalues of $C_n(\bell;\vec{a})$ in the subspace transforming according to $\cS_j$ excluding any eigenvalues in the Dirichlet spectrum. Note that eigenfunctions transforming according to the irreducible representation $\cS_j$ which vanish at all vertices of the circulant graph correspond to eigenfunctions on the quotient graph which vanish at the central vertex, $\phi=0$, in which case $\sin k \ell_h = 0$ for some $h$ and $k = m\pi / \ell_h$. So $k^2$ is in the Dirichlet spectrum of the circulant graph. To obtain the whole spectrum we must therefore take the union over $j = 0,\dots,n-1$ and include any eigenvalues in the Dirichlet spectrum. To take the union over the representations we can multiply the secular equations obtaining theorem \ref{thm:symseculareq}.

\subsection{Dirichlet spectrum}\label{sec:dirchlet}

For a metric graph $C_n(\vec{L};\vec{a})$ the secular equation given in theorem \ref{thm:seculareq} only identifies values of $k$ corresponding to eigenvalues when $k$ is not in the Dirichlet spectrum of the graph. However, given a graph where the $k$-spectrum and Dirichlet spectrum overlap this can be removed by perturbing the edge lengths $\vec{L}$.

Because of the additional symmetry of the graph $C_n(\bell;\vec{a})$, eigenvalues in the Dirichlet spectrum also appear with Neumann vertex conditions. To determine which Dirichlet eigenvalues occur and with what multiplicity, it is useful to return to the quotient graph.

Consider a value of $k$ in the Dirichlet spectrum.  Then $k = m\pi / \ell_g$ for some given value of $g \in \{1,\dots,d\}$ and some fixed $m\in\NN$.  For $h\neq g$, solutions to \eref{schrodingeredge} on the edge $h$ of the quotient graph are given by the piecewise solution \eref{piecewisesoln}. However, on the edge $g$, we have the solution
\begin{equation}\label{dirichletpiecewisesoln}
\fl
\psi_g(x) = \cases{
            \phi \cos\left(\frac{m\pi}{\ell_g} x\right) + \beta_g \sin\left(\frac{m\pi}{\ell_g} x\right) & if $0 \leq x < \ell_g / 2$ \\
            \phi \rme^{-\rmi a_g \theta_j} \cos\left(\frac{m\pi}{\ell_g} x\right) + \beta_g \rme^{-\rmi a_g \theta_j} \sin\left(\frac{m\pi}{\ell_g} x\right) & if $\ell_g / 2 < x \leq \ell_g$
            }
\end{equation}
where $\phi$ is the value of the eigenfunction of the quotient graph at the central vertex.

By continuity, $\psi_g(\ell_g) = \phi$. Evaluating $\psi_g(x)$ at $x = \ell_g$, we obtain the condition
\begin{equation}\label{dirichletcondition1}
\phi \rme^{-\rmi a_g \theta_j} (-1)^m = \phi \ .
\end{equation}
Clearly this is satisfied if $\phi = 0$, which yields the quotient graph eigenfunction
\begin{equation}
\psi_g(x) = \cases{
            \beta_g \sin\left(\frac{m\pi}{\ell_g} x\right) & if $0 \leq x < \ell_g / 2$ \\
            \beta_g \rme^{-\rmi a_g \theta_j} \sin\left(\frac{m\pi}{\ell_g} x\right) & if $\ell_g / 2 < x \leq \ell_g$
            }
\end{equation}
on the edge set $g$ and $\psi_h(x) = 0$ for $h\neq g$. If we apply the Neumann condition \eref{quotientvertexcondition2} at the central vertex we see that, for a nontrivial solution,
\begin{equation}
\rme^{\rmi a_g \theta_j} = (-1)^m  \ .
\end{equation}
For $m$ even this requires $j a_h$ to be a multiple of $n$ and for $m$ odd $2 j a_h$ must be an odd multiple of $n$. 

Returning to \eref{dirichletcondition1} and supposing that $\phi \neq 0$, we see that $\rme^{\rmi a_g \theta_j} = (-1)^m$ is now required to satisfy continuity at the central vertex. Substituting in the Neumann condition \eref{quotientvertexcondition2} with $\psi'_g (0) = \psi'_g(\ell_g)$ we obtain,
\begin{equation}\label{unsatisfiedcondition}
\sum_{\substack{h=1 \cr h\neq g}}^d  \cos(\theta_j a_h) \csc\left(\frac{m\pi\ell_h}{\ell_g}\right) - \cot\left(\frac{m\pi\ell_h}{\ell_g}\right) = 0 \ .
\end{equation}

The derivative of the left hand side of \eref{unsatisfiedcondition} with respect to an edge length is non-zero. Hence if this condition is satisfied for some set of edge lengths a perturbation of any edge length will remove this.  If we consider choosing each of the edge lengths $\ell_h \in (1-\epsilon,1+\epsilon)$, then for almost all edge lengths, \eref{unsatisfiedcondition} is not satisfied and the only Dirichlet eigenvalues correspond to eigenfunctions which vanish at the central vertex. This leads to the following lemma.

\begin{lem}\label{lem:dirichlet}
  Let $C_n(\bell;\vec{a})$ be a quantum circulant graph ($d \geq 2$) with symmetric edge lengths, equipped with the Laplace operator and satisfying Neumann vertex conditions. Then for almost every $\bell \in (1-\epsilon,1+\epsilon)^d$ and all $m\in\NN$, $m^2\pi^2 / \ell_g^2$ is in the spectrum of the graph with multiplicity $|J|$ where,
  \begin{equation}\label{dirichletcondition2}
  \fl
  J = \left\{j\in \{0,\dots,n-1\} : 2ja_g = q n \textrm{ for some odd/even } q \textrm{ when } m \textrm{ is odd/even} \right\} \ ,
  \end{equation}
  and the corresponding eigenfunctions are zero at the graph vertices.
\end{lem}
Consequently, when $n$ is odd, $m^2\pi^2 / \ell_g^2$ is not in the spectrum for odd values of $m$ and, for any $n$, $m^2\pi^2 / \ell_g^2$ is always in the spectrum of the graph for even $m$.

\subsection{Weyl Law}

For a general quantum graph one can define an \emph{eigenvalue counting function} $N(a,b)$; the number of eigenvalues $\lambda=k^2$ with $\sqrt{\lambda}=k$ in the interval $(a,b)$.  
It is sensible to count the number of eigenvalues in terms of $\sqrt{\lambda}=k$ as then $N(a,b)$ grows linearly with the length of the interval.   Such a result is referred to as a \emph{Weyl law}, see e.g. \cite{BK13}.
\begin{lem}\label{lem:Weyl law}  For a quantum graph with standard (or Neumann-like) conditions at every vertex,
\begin{equation}
N(a,b)=\frac{(b-a)\cL}{\pi} + O(1) \ ,
\end{equation}
where $\cL=\sum_{(i,j)\in \cE} L_{i,j}$ is the total length of the graph and the remainder term is bounded above and below by constants independent of $k$.
\end{lem} 
The result can be proved comparing the graph to the set of disjoint intervals with Dirichlet conditions at the end points by changing a standard condition at a vertex to a Dirichlet condition, vertex by vertex. 

For symmetric circulant graphs the secular equation, theorem \ref{thm:symseculareq}, for eigenvalues that are not in the Dirichlet spectrum, and lemma \ref{lem:dirichlet}, for eigenvalues in the Dirichlet spectrum, must be consistent with the Weyl law, lemma \ref{lem:Weyl law}.  As a first application of the symmetric circulant graph results it is interesting to see this directly.

Looking at theorem \ref{thm:symseculareq} we can examine each function $p_j(k)$ individually to count zeros in $(a,b)$ which correspond to eigenvalues.  As $p_j(k)$ is strictly increasing there is exactly one zero of $p_j(k)$ between each pair of adjacent asymptotes.  Consequently to count zeros of $p_j(k)$ in $(a,b)$ one can count asymptotes.  

Let $N_j(a,b)$ be the number of zeros of $p_j(k)$ in the interval $(a,b)$.   Consider $p_0(k)$, the spacing between asymptotes of $\tan(k \ell_h / 2)$ is $ 2\pi / \ell_h $, so
\begin{equation}\label{rootestimateg0}
\frac{(b-a)\cL}{2\pi n}-d-1 \leq N_0(a,b) \leq \frac{(b-a)\cL}{2\pi n} +d+ 1 \ ,
\end{equation}
where $\cL = n \sum_{h=1}^d \ell_h $ is the total length of the graph and the set of edge lengths 
$\{\ell_1,\dots , \ell_d\}$ is incommensurate. 

For $p_{n/2}(k)$, which only occurs when $n$ is even, the asymptotes are $(2m-1)\pi / \ell_h$ for even $a_h$ and $2m\pi / \ell_h$ for odd $a_h$. In either case, the spacing between asymptotes is $2\pi / \ell_h $ and so \eref{rootestimateg0} also bounds the number of zeros of $p_{n/2}(k)$.

For $1 \leq j \leq \lf (n-1) / 2 \rf $, the asymptotes of $p_j(k)$ will be zeros of $\sin(k\ell_h)$, unless
\begin{equation} \label{algebraicDirichletcondition}
2 j a_h = 0 \Mod n \ ,
\end{equation}
which is equivalent to the condition to see additional Dirichlet eigenvalues associated with this value of $j$, see lemma \ref{lem:dirichlet}. Consequently, every missing asymptote corresponds to an additional Dirichlet eigenvalue. Let
\begin{equation}
J_h = \left\{j\in \left\{1,\dots,\lf \frac{n-1}{2} \rf \right\} : 2 j a_h  = 0 \Mod n \right\} \ . 
\end{equation}
Then, if $j\notin J_h$ for any $h=1,\dots,d$,
\begin{equation}\label{rootestimatepJ1}
\frac{(b-a)\cL}{\pi n}- d -1 \leq N_j(a,b) \leq \frac{(b-a)\cL}{\pi n} +d+ 1 \ .
\end{equation}
And if we replace the number of roots of $p_j$ with the number of eigenvalues associated with $j$, including Dirichlet eigenvalues satisfying the algebraic condition \eref{algebraicDirichletcondition}, then for any $j$,   
\begin{equation}\label{eigenestimatepJ1}
\frac{(b-a)\cL}{\pi n} -d-1 \leq \textrm{\# eigenvalues} \leq \frac{(b-a)\cL}{\pi n} +d+ 1 \ .
\end{equation}

To obtain a Weyl law for edge-symmetric circulant graphs we must sum the eigenvalues associated to each $j$, noting that for $1 \leq j \leq \lf (n-1) / 2 \rf$ each eigenvalue has multiplicity two. Every graph also has the Dirichlet eigenvalues $2m\pi / \ell_h$ for $m \in \NN$ and when $n$ is even $(2m+1)\pi / \ell_h$ is also in the Dirichlet spectrum along with the roots of $p_{n/2}(k)$. Then putting the bounds on the eigenvalues for each $j$ together with the Dirichlet eigenvalues, we obtain,
%
%

\begin{cor}
  For a quantum circulant graph $C_n(\bell;\vec{a})$ with symmetric edge lengths the number of eigenvalues in the interval $(a,b)$ is
  \begin{equation}\label{totalcount}
  N(a,b) = \frac{(b-a)\cL}{\pi} + R
  \end{equation}
  where $\cL = n \sum_{h=1}^d \ell_h $ is the total length of the graph, and $\lv R \rv \leq nd + n + d $, so it is uniformly bounded in the length of the interval.
\end{cor}
This result shows theorem \ref{thm:symseculareq} and lemma \ref{lem:dirichlet} are consistent with the Weyl law.  It should be noted that, the typical argument used to prove lemma \ref{lem:Weyl law} produces a slightly tighter bound, $|R|\leq nd+n$ the sum of the number of edges and vertices \cite{B17},  see also \cite{BK13} for further discussion.

\section{Intermediate statistics}\label{sec:intermediate}

To analyze the spectral statistics of the quantum circulant graph with symmetric edge lengths we should consider the subspectra  transforming according to each irreducible representation $\cS_j$ individually \cite{KR97}. A secular equation \eref{eq:quotientsecular} for a single subspectra is closely related to the secular equation of the Dirac operator on a rose graph investigated in \cite{HW12}.  A \emph{rose graph} is a single vertex with $d$ edges that are loops which start and end at the vertex, so the rose has the topology of the quotient graph, figure \ref{fig:dummyvertices}(b).

In \cite{HW12} eigenvalues of the Dirac operator on the rose are values of $k$ which solve a secular equation,
\begin{equation}\label{eq:Diracsecular}
\sum_{h=1}^d \cos(\theta_h) \csc(k\ell_h) - \cot(k\ell_h) = 0 \ ,
\end{equation}
where $\ell_h$ is the length of one of the edges and $\rme^{\pm\rmi\theta_h}$ are the eigenvalues of an $\mathrm{SU}(2)$ matrix corresponding to a spin rotation along the edge with length $\ell_h$. For a spectrum $\{\lambda_j\}$, scaled so that the mean spacing is one, the two-point correlation function $R_2(x)$ is defined by,
\begin{equation}\label{eq:R2defn}
\lim_{N\to\infty} \frac{1}{N} \sum_{i=1}^N \sum_{j=1}^N g(\lambda_i - \lambda_j) = g(0) + \int_{-\infty}^\infty g(x) R_2(x) \dif x \ .
\end{equation}
The function $g$ is a suitably defined test function.  If $g$ approximates the characteristic function of an interval we see $R_2$ is a measure of the pairs of eigenvalues whose separation falls in the given interval.  

The two-point correlation function of Neumann star graphs and \v Seba billiards have been analyzed in detail in \cite{BGS99,BGS01sing}, where they show that as $x\to 0$,
\begin{equation}
R_2(x) \sim \frac{\pi \sqrt{3}}{2} x \ .
\end{equation}
For $x\to\infty$ a full series expansion has been derived \cite{BBK01,BK99},
\begin{equation}\label{largeR2star}
R_2(x) \sim 1+ \frac{2}{\pi^2 x^2} +\frac{76}{\pi^4 x^4} + \Or\left(\frac{1}{x^{6}}\right) \ .
\end{equation}

The methods employed for the star graph were developed for a Dirac operator with symplectic symmetry on the rose in \cite{HW12}.  For a graph where the spin transformations are chosen randomly from $\mathrm{SU}(2)$ with Harr measure, as $x\to 0$,
\begin{equation}
R_2(x) \sim \frac{\pi c}{6} x \ ,
\end{equation}
where $c\approx 6.781$ and 
for $x\to\infty$,
\begin{equation}\label{largeR2Diracrose}
R_2(x) \sim 1+ \frac{2}{\pi^2 x^2} -\frac{13}{8\pi^4 x^4} + \Or\left(\frac{1}{x^{6}}\right)  \ .
\end{equation}
Significantly neither the distributions for the Laplacian on a star graph or the Dirac operator on a rose graph correspond to the random matrix distribution typically seen for the spectral statistics of graph models but are rather examples of intermediate statistics, statistics intermediate between those of the random matrix ensembles and the Poisson distribution associated with integrable classical dynamics. Systems with intermediate spectral statistics exhibit properties such as linear level repulsion and an exponential decay in the probability of large spacings. Examples that have been investigated include, the Anderson model at the metal-insulator transition point \cite{SSSetal93}, Aharonov-Bohm integrable billiards \cite{BGS01cmp,DJM94, RF01}, \v{S}eba billiards \cite{AS91,S90,SZ91} and polygonal billiards with rational angles \cite{BGS99,GJ98, PJ96}, eigenphases of quantum maps \cite{GMK04} and a one dimensional gas interacting with a logarithmic potential \cite{BGS01intr}.

\subsection{Small parameter behavior of the two-point correlation function}

We apply the analysis from \cite{BGS99,BGS01sing} to obtain the small parameter asymptotics of the two-point correlation function. The method was applied to the Dirac operator on a rose graph in \cite{HW12}.  We reproduce the essentials of the argument here for completeness. Given a random real meromorphic function with poles on the real axis the statistics of small spacings of zeros are approximated by zeros of a function with three randomly distributed poles,
\begin{equation}\label{randomfunction}
\frac{r_1}{k-c_1} + \frac{r_2}{k-c_2} + \frac{r_3}{k-c_3} = 0 \ .
\end{equation}
$r_1,r_2,r_3$ are random residues and $c_1,c_2,c_3$ are random locations of the poles.

Rearranging \eref{randomfunction},
\begin{equation}
\fl
(r_1 + r_2 + r_3)k^2 - (r_1(c_2+c_3) + r_2(c_1+c_3) + r_3(c_1+c_2))k + r_1c_2c_3 + r_2c_1c_3 + r_3c_1c_2 = 0
\end{equation}
whose solutions are
\begin{equation}
k_{\pm} = \frac{r_1(c_2+c_3) + r_2(c_1+c_3) + r_3(c_1+c_2) \pm \sqrt{\cD}}{2(r_1 + r_2 + r_3)} \ ,
\end{equation}
where
\begin{equation}\label{discriminant1}
\fl
\cD = (r_1(c_2+c_3) + r_2(c_1+c_3) + r_3(c_1+c_2))^2 - 4(r_1c_2c_3 + r_2c_1c_3 + r_3c_1c_2)(r_1 + r_2 + r_3) \ .
\end{equation}
%
%
%
%
%
The leading contribution to the two-point correlation function is obtained by averaging over the positions of the poles and the random residues,
\begin{equation}\label{expectedvalue1}
\EE(R_2(x)) \approx \frac{1}{2} \EE\left\{\delta(x - \Delta k)\right\} \ ,
\end{equation}
where
\begin{equation}
\Delta k = k_{+} - k_{-} = \frac{\sqrt{\cD}}{r_1 + r_2 + r_3} \ .
\end{equation}
Carrying out the average over the position of the poles one finds,
\begin{equation}\label{eq:expectationofresidues}
\EE(R_2(x)) \approx \frac{\pi x}{6} \EE\left\{\sqrt{\frac{(r_1 + r_2 + r_3)^3}{r_1 r_2 r_3}}\right\} \ .
\end{equation}

Noting,
\begin{equation}
\fl
\cos\theta \csc k - \cot k = \sum_{m=-\infty}^\infty ((-1)^m \cos\theta -1) \left(\frac{1}{k+\pi m} - \frac{m\pi}{1+m^2\pi^2}\right)
\end{equation}
we see that the secular equations \eref{eq:quotientsecular} and \eref{eq:Diracsecular} have pole expansions with poles at $k = m\pi / \ell_h$ for $m \in \mathbb{Z}$ and $h = 1,\dots,d$ and residues $r_i = (-1)^{m_i} \cos\theta_i - 1$.  For poles to be closely spaced, they must be closely spaced for different edges $h$.  In the Dirac rose model $\theta_i$ is an eigenphase of an $\mathrm{SU}(2)$ matrix selected randomly with Harr measure and the expectation on the right hand side of \eref{eq:expectationofresidues} can be evaluated by integrating over three independent random phases.  However, for an edge-symmetric circulant graph $\theta_i$ is one of a finite number of  phases  
$\theta_{i} \in \{2\pi j a_h / n : h = 1,\dots,d\}$ depending on the edge $h$ the associated residue comes from and the representation of the cyclic group, determined by $j=1,\dots,n-1$.  If we generate the circulant graph randomly, so each $a_h\in \{1,\dots,\lf (n-1) / 2 \rf\}$ is present with probability $p$ where $0<p<1$, we can consider the $\theta_i$ to be random variables.

To describe the statistics of a typical subspectrum of an edge-symmetric circulant graph we want to carry out the average in \eref{eq:expectationofresidues} with
\begin{equation}
r_i = 1 - \cos\theta_{i}
\end{equation}
where the $\theta_{i}$ are randomly chosen $a_h$ multiples of $2\pi j/n$.  To make a general prediction for a class of small parameter asymptotics we assume that $n$ is a large prime, so $j$ and $n$ have no common factors.  Then $a_h$ multiples of $2\pi j/n$ are evenly distributed in $[0,\pi]$ and as $n$ is large we can model $\theta_i$ with independent uniformly distributed random numbers in $[0,\pi]$. 

However, the expression \eref{eq:expectationofresidues} diverges if the probability of small $r$ is large, which is the case here. The divergence of \eref{eq:expectationofresidues} was dealt with for random residues in \cite{BGS01sing}. Following their analysis, the leading behavior for small $x$ is,
\begin{equation}
\EE\left\{\sqrt{\frac{(r_1 + r_2 + r_3)^3}{r_1 r_2 r_3}} \right\} \to 3\EE\left\{r\right\}\EE\left\{\frac{1}{\sqrt{r}}\right\}^2 \ .
\end{equation}
Then 
\begin{equation}
\EE\left\{r\right\} = \frac{1}{\pi} \int_{0}^{\pi} (1 - \cos\phi) \dif\phi = 1
\end{equation}
and the divergent behavior comes from $\EE\{1 / \sqrt{r}\}$. Introducing a cutoff $\phi_0$,
\begin{equation}
\EE\left\{\frac{1}{\sqrt{r}}\right\} = \frac{1}{\pi} \int_{\phi_0}^{\pi} \frac{\dif\phi}{\sqrt{1 - \cos\phi}}  \approx \frac{\sqrt{2}}{\pi} \ln\phi_0 \ .
\end{equation}
With logarithmic accuracy, $\phi_0$ is proportional to $x$, $\phi_0 \to x/c $.  So as $x$ approaches $0$,
\begin{equation}\label{smallxapprox}
\EE(R_2(x)) \approx \frac{1}{\pi} \ln^2\left(\frac{x}{c}\right) x \ .
\end{equation}

\subsection{Large parameter behavior of the two-point correlation function}

We apply the method used to analyze the large $x$ behavior of the two-point correlation function of the Dirac operator on a rose graph in \cite{HW12} to the subspectra of the circulant graph with symmetric edge lengths. In order to find asymptotics of the two-point correlation function, we consider its Fourier transform, known as the \emph{form factor}.  The small parameter asymptotic of the form factor will determine the large parameter behavior of the two-point correlation function. For a compact quantum graph with incommensurate edge lengths over the rationals, the form factor can be expressed as a sum over pairs of periodic orbits on the graph \cite{BK13},
\begin{equation}\label{formfactor}
K(\tau) = \frac{1}{4\cL^2} \sum_{p,q\in\cP} \frac{A_p A_q L_p L_q}{r_p r_q} \delta\left(\tau - \frac{L_p}{2\cL}\right) \delta_{L_p, L_q} \ ,
\end{equation}
where $p$ and $q$ are periodic orbits; equivalence classes of closed paths modulo cyclic shifts. $\cP$ is the set of all periodic orbits on the graph. A periodic orbit $p\in\cP$ of $m$ edges can be written as a sequence of adjacent vertices, $p=(v_1,\dots,v_m)$ with $v_i\in\cV$, and $v_i $ connected to  $v_{i+1}$ by an edge $e_i\in \cE$ for $i=1,\dots,m-1$,  $v_m$ is connected to  $v_1$ by $e_m$.  Associated to an orbit $p$ we have the metric length of the orbit $L_p$, the sum of the lengths of the edges traversed in the orbit.  An orbit $p$ can be a repetition of a shorter orbit, the repetition number $r_p$ is the maximum number of repetitions of a shorter orbit contained in $p$. $A_p$ is the stability amplitude of the orbit $p$ the product of the elements of the scattering matrices associated to the vertices around the orbit.  $\cL$ is the total length of the graph.

For the quotient graph, defined in section \ref{sec:quotientgraph},  we have two types of vertices. The central vertex has standard vertex matching conditions, \eref{vertexcondition1} and \eref{vertexcondition2}. Scattering amplitudes that relate incoming to outgoing plane waves at the central vertex are $(1/d)-1$ 
 for back scattering onto the edge the wave came from and $1/d$ for scattering onto other edges.   Passing through the degree two vertices of the quotient graph plane waves pick up a scattering amplitude $\rme^{\pm \rmi a_h \theta_j}$ with no back scattering.  To approximate the averaged form factor we will evaluate the contribution of those orbits which backscatter at the central vertex the maximum number of times.  In the limit $d \to \infty$ backscattering is increasingly heavily weighted relative to scattering onto a new edge and provides a good approximation to the form-factor of Neumann star graphs \cite{BK99}. 

Let $\gL$ be the set of all possible lengths of periodic orbits. The sum in \eref{formfactor} can be arranged to sum over \emph{degeneracy class} of orbits with the same length.
\begin{equation}
K(\tau) = \frac{1}{4\cL^2} \sum_{L\in\gL} L^2 \delta\left(\tau - \frac{L}{2\cL}\right) \left(\sum_{\substack{p\in\cP \cr L_p=L}} \frac{A_p}{r_p}\right)^2 
\end{equation}
Define,
\begin{equation}
\tilde{K}(t,d) = \frac{d}{2\cL^2} \sum_{L\in\gL} L^2 \left(\sum_{\substack{p\in\cP_t \cr L_p=L}} \frac{A_p}{r_p}\right)^2 \ ,
\end{equation}
where $\cP_t$ is the set of periodic orbits traversing $t$ edges.  As there is no back-scattering at the vertices of degree two in the quotient graph, to simplify the presentation, we here regard each loop of the quotient graph as a single edge of length $\ell_h$, rather than two edges of length $\ell_h / 2$. Then $\tilde{K}(t,B)\to K(\tau)$ weakly as $d\to\infty$ when $t / 2d \to\tau$ as $d\to\infty$.

%
\begin{equation}\label{expectedvalueformfactor}
\EE\left(K(\tau)\right) = \lim_{\substack{d\to\infty \cr t/2d\to\tau}} \sum_{j=1}^d \tilde{K}_j(t,d) \ ,
\end{equation}
where
\begin{equation}\label{Ktildej}
\tilde{K}_j(t,d) = \frac{d}{2\cL^2} \sum_{\substack{L\textrm{ restricted} \cr \textrm{to } j\textrm{ edges}}} L^2 \EE \left(\sum_{\substack{p\in\cP_t \cr L_p=L}} \frac{A_p}{r_p}\right)^2 \ .
\end{equation}
The sum over $L$ has been ordered according to the number of distinct loops $j$ of the quotient graph that the orbit is confined to.

We first consider the special case when $j=1$, where the periodic orbits remain on a single loop of the quotient graph.  As in the case of the Dirac operator the result depends on whether the lengths of the orbits are odd or even.


When $t$ is even, for some fixed edge (loop) $e$, periodic orbits with maximum back-scattering will have the form
\begin{equation}
p = e \bar{e} e \bar{e} \cdots e \bar{e} e \bar{e} 
\end{equation}
that is $t / 2$ repetitions of $e\bar{e}$, where $\bar{e}$ is the reversal of $e$. Scattering from $e$ onto $\bar{e}$, the scattering coefficient is $\left(1/d - 1\right)$, multiplied by a phase factor $\omega = \rme^{\pm\rmi\theta_{e}}$ picked up at the dummy vertex at the halfway point along the edge. However, as we scatter from $\bar{e}$ to $e$, the scattering coefficient is $\left(1/d - 1\right)$, but the additional phase factor is $\omega^{-1} = \rme^{\mp\rmi\theta_{e}}$. Thus the product of the scattering amplitudes for a periodic orbit of topological length $t$ is
\begin{equation}
A_p = \left(\frac{1}{d} - 1\right)^t \ .
\end{equation}
The repetition number for such an orbit is $r_p = t / 2$ and there are $d$ choices for the edge $e$. Combining this information and approximating the metric length of the orbit by $t$ and the length of the graph by $d$ for a set of edge lengths in an interval $(1-1/d,1+1/d)$,
\begin{equation}
\tilde{K}_1(t,d) \approx 
2 \left(1 - \frac{1}{d}\right)^{2t} \ .
\end{equation}
Then,
\begin{equation}\label{K1even}
\lim_{\substack{d\to\infty \cr t/2d\to\tau}} \tilde{K}_1(t,d) \approx \lim_{d\to\infty} 2\left(1 - \frac{1}{d}\right)^{4d\tau} = 2\rme^{-4\tau} \ .
\end{equation}

In the case where $t$ is odd, there are two orbits with maximal back-scattering for a fixed edge $e$,
\begin{equation}
p = \cases{
    e \bar{e} e \bar{e} \cdots e \bar{e} \bar{e} \\
    \bar{e} e \bar{e} e \cdots \bar{e} e e
    } \ .
\end{equation}
In both of these cases, the orbit will pick up $t-1$ scattering coefficients of the form $\left(1/d - 1\right)$, but when scattering from $e$ to $e$ or $\bar{e}$ to $\bar{e}$, the scattering amplitude is $1/d$ with an additional phase factor $\omega$ or $\omega^{-1}$ respectively. For both orbits, $r_p = 1$, so
\begin{equation}
\tilde{K}_1(t,d) \approx \frac{t^2}{2d^2} \left(1 - \frac{1}{d}\right)^{2t-2} \EE\left\{(\omega + \omega^{-1})^2\right\} \ .
\end{equation}
Consequently,
\begin{eqnarray}
\lim_{\substack{d\to\infty \cr t/2d\to\tau}} \tilde{K}_1(t,d) &\approx \lim_{d\to\infty} 2\tau^2 \left(1 - \frac{1}{d}\right)^{4d\tau - 2} \EE\left\{4\cos^2 \theta_{e}\right\} \\
&= 4 \tau^2 \rme^{-4\tau} \ , \label{K1odd}
\end{eqnarray}
where we will see that $\EE\left\{\cos^2 \theta_{e} \right\} = 1/2$ (this is not surprising given the argument in the previous section but we postpone the discussion to the end).
Since the relative density of odd and even values for $t$ is $1/2$, the total contribution from orbits confined to a single loop is an average of formulas \eref{K1even} and \eref{K1odd}; $(1 + 2\tau^2)\rme^{-4\tau}$.

In general, if a periodic orbit is restricted to $j$ loops of the quotient graph and has maximal back-scattering it will remain on a single edge $e_l$ bouncing back and forth $t_l$ times before scattering to a new edge $e_{l+1}$, with
\begin{equation}\label{decomposet}
t_1 + \cdots + t_j = t \ .
\end{equation}
In order to weight each contribution, we first consider the relative density of odd versus even $t_l$'s. As each $t_l$ can be odd or even, the contribution of a given combination picks up a weight $1 / 2^j$. Then, to decompose $t$ into the sum \eref{decomposet} we must choose $j-1$ transition points from the $t-1$ possibilities. Although the choice of transition points and the selection of odd verses even $t_l$ are not independent the odd and even values of $t_l$ are evenly distributed, and so for large $t$ we approximate the weight of each contribution by the product,
\begin{equation}
\frac{1}{2^j} {t-1\choose j-1} \sim \frac{t^{j-1}}{2^j(j-1)!} \ .
\end{equation}

When back-scattering from an edge onto its reversal, the scattering coefficient is $(1/d) - 1$ and scattering to a new edge the coefficient is $1/d$, as before. 
Since we have $j$ edge transitions, the product of the scattering amplitudes at the central vertex is,
\begin{equation*}
\left(\frac{1}{d} - 1\right)^{t-j} \frac{1}{d^j} \ .
\end{equation*}
There is also a phase associated with the orbit which is picked up at the degree two vertices of each loop. When $t_l$ is even the product of the phases cancel, so we must only consider those $t_l$ which are odd. If there are $r$ indices $\left\{i_1,\dots,i_r\right\}$ for which $t_{i_l}$ is odd, this product will be
\begin{equation*}
\omega_{e_{i_1}}^{\alpha_{i_1}} \cdots \omega_{e_{i_r}}^{\alpha_{i_r}} \ ,
\end{equation*}
where $\omega_{e_{i_l}} = \exp\left(\rmi\theta_{e_{i_l}}\right)$ and $\alpha_{i_l} = \pm 1$. Since there are $2^j$ members of each degeneracy class and $2^r$ combinations of $\alpha_{i_l} = \pm 1$, we must evaluate,
%
%
\begin{eqnarray}
\fl
\EE\left\{ \left(2^{j-r} \sum_{\alpha_{i_l}=\pm 1} \left(\omega_{e_{i_1}}^{\alpha_{i_1}} \cdots \omega_{e_{i_r}}^{\alpha_{i_r}} \right) \right)^2 \right\} &= 2^{2(j-r)} \EE\left\{2^{2r} \cos^2\theta_{e_{i_1}} \cdots \cos^2\theta_{e_{i_r}} \right\} \\
&= 2^{2j-r} \ .
\end{eqnarray}
Finally, each orbit has repetition number $r_p = 1$, there are $j\choose r$ ways of choosing $r$ odd indices, and the number of choices of a set of $j$ distinct edges is,
\begin{equation}
\frac{1}{j} \frac{d!}{(d-j)!} \sim \frac{d^j}{j} \ .
\end{equation}
%
%
%
%
%
%
%
%
%
%
%
%
%
%
%
%
Substituting in the definition of $\tilde{K}_j$ from \eref{Ktildej},
\begin{equation}
\tilde{K}_j(t,d) \approx \frac{t^{j+1}}{(2d)^{j+1} j!} \left(1 - \frac{1}{d}\right)^{2t-2j} \left\{\sum_{r=0}^{j} {j\choose r} 2^{2j-r} \right\} 
\end{equation}
Then taking $d\to\infty$ with $t/2d \to \tau$,
\begin{eqnarray}
\lim_{\substack{d\to\infty \cr t/2d\to\tau}} \tilde{K}_j(t,d) &\approx \lim_{d\to\infty} \frac{\tau^{j+1}}{j!} \left(1 - \frac{1}{d}\right)^{4d\tau-2j} \left\{6^j\right\} \\
&= \tau^{j+1} \rme^{-4\tau} \frac{6^j}{j!} \ .
\end{eqnarray}

Substituting these results in the expansion \eref{expectedvalueformfactor},
\begin{eqnarray}
\EE\left(K(\tau)\right) &\approx (1 + 2\tau^2)\rme^{-4\tau} + \tau \rme^{-4\tau} \sum_{j=2}^\infty \frac{6^j}{j!} \tau^j \\
& = \left(1 - \tau - 4\tau^2 \right)\rme^{-4\tau} + \tau \rme^{2\tau} \ .
\end{eqnarray}
And as a Maclaurin series,
\begin{equation}\label{formfactormaclaurin}
\EE\left(K(\tau)\right) \approx 1 - 4\tau + 10\tau^2 - \frac{2}{3}\tau^3 - \frac{28}{3}\tau^4 + \Or(\tau^5) \ .
\end{equation}

Since the form factor is the Fourier transform of the two-point correlation function, small $\tau$ asymptotics for $K(\tau)$ determine the large $x$ asymptotics of $R_2(x)$. In particular, if $k(\tau)$ and $r(x)$ are related via
\begin{equation}
1 - k(\tau) = \int_{-\infty}^\infty (1 - r(x))\rme^{2\pi\rmi x \tau} \dif x \ ,
\end{equation}
then for $k(\tau)$ even and 
\begin{equation}
k(\tau) \sim 1 + \sum_{l=1}^\infty a_l \tau^l
\end{equation}
we have
\begin{equation}
r(x) \sim 1 + 2\Re\left\{\sum_{l=1}^\infty \left(\frac{-\rmi}{2\pi}\right)^{l+1} \frac{a_l l!}{x^{l+1}}\right\} \ .
\end{equation}
Using the coefficients of the Maclaurin series \eref{formfactormaclaurin} yields the following expansion for $R_2(x)$ when $x$ is large,
\begin{equation}\label{largexapprox}
\EE(R_2(x)) \sim 1 + \frac{2}{\pi^2 x^2} - \frac{1}{2\pi^4 x^4} + \Or\left(\frac{1}{x^6}\right) \ .
\end{equation}
This result can be compared with the large $x$ asymptotic for the two-point correlation functions of the star graph \eref{largeR2star} and the Dirac operator on a rose graph \eref{largeR2Diracrose}.  We see that each case differs from the others only from the coefficient of $x^{-4}$ in the expansion, which is $76 / \pi^4$ for the star and $-13 / 8\pi^4$ for the Dirac rose.  

\subsection{Expectation value of $\cos^2 \theta_e$}

The argument in the previous section used the expectation value,
\begin{equation}\label{expectation value}
\EE\left\{\cos^2 \theta_{e} \right\} = \frac{1}{2} \ .
\end{equation} 
This compares with an expectation value $\EE\left\{\cos^2 \theta \right\} = 1$ for $\theta$ the eigenphase of an $\mathrm{SU}(2)$ matrix distributed with Haar measure, which appears in the corresponding result for the Dirac operator on a rose graph \cite{HW12}. 

For a quantum circulant graph $\theta_e=2\pi j a_e/n$ where the $a_e$ are chosen randomly from 
$\{1,\dots, \lf n /2 \rf \} $, each appearing with probability $p$.   Then,
\begin{eqnarray}\label{expectation value 2}
\EE\left\{\cos^2 \theta_{e} \right\} &= \frac{1}{ \lf n /2 \rf} \sum_{l=1}^{ \lf n /2 \rf} \cos^2 \left( \frac{2\pi j l }{n} \right) \\
&= \left\{ \begin{array}{lcl} 
\frac{1}{ n} \sum_{l=1}^{ n} \cos^2 \left( \frac{2\pi j l }{n} \right)  && $n$ \textrm{ even} \\
 \frac{1}{ n-1} \sum_{l=1}^{ n-1} \cos^2 \left( \frac{2\pi j l }{n} \right)  && $n$ \textrm{ odd} \\
 \end{array} \right. \ . 
\end{eqnarray}

The following lemma follows from the orthogonality of the characters of $\mathbb{Z}_n$.
\begin{lem}
For $j=0,\dots, n-1$,
\begin{equation}
\sum_{l=1}^{n} \cos^2 \left(  \frac{2\pi j l }{n}\right) 
 = \left\{ \begin{array}{lcl}
n & & \textrm{if } j=0 \textrm{ or } j=\frac{n}{2} \\
\frac{n}{2} & & \textrm{otherwise} \\
\end{array} \right. \ .
\end{equation}
\end{lem}
Consequently for $n$ even,
\begin{equation}\label{expectation value 3}
\EE\left\{\cos^2 \theta_{e} \right\} 
= \left\{ \begin{array}{lcl} 
1  && \textrm{if } j=\frac{n}{2}  \\
 \frac{1}{ 2}   && \textrm{otherwise} \\
 \end{array} \right. \ ,
\end{equation}
and if $n$ is odd,
\begin{equation}\label{expectation value 4}
\EE\left\{\cos^2 \theta_{e} \right\} 
=  \frac{1}{ 2} \left( 1- \frac{1}{n-1}\right)  \ .
\end{equation}
Then, in the limit of large graphs and for representations $j\neq 0,  n/2$, we see that the required result for the expectation value of $\cos^2 \theta_e$ holds.

For the representations $j=0$ and $j=n/2$, and using the alternative expectation value from  (\ref{expectation value}),
\begin{eqnarray}
\EE\left(K(\tau)\right) 
&\approx  \left(1 - \tau - 4\tau^2 \right)\rme^{-4\tau} + \tau \rme^{4\tau}  \\
&\approx  1 - 4\tau + 12\tau^2 + \frac{16}{3}\tau^3 + \Or(\tau^5) \ .
\end{eqnarray}
Subspectra transforming according to these irreducible representations have large $x$ asymptotics,
\begin{equation}
\EE(R_2(x)) \sim 1 + \frac{2}{\pi^2 x^2} + \frac{4}{\pi^4 x^4} + \Or\left(\frac{1}{x^6}\right) \ .
\end{equation}
These two representations, the trivial representation $j=0$ and the representation $j=n/2$, are also the two representations for which the secular equation $p_j(k)=0$ takes a different form, see (\ref{pj}).

\section{Numerical results}\label{sec:numerics}

\begin{figure}[htb!]
\centering
  \subfloat[][]{
  \centering
   \includegraphics[width=6.5cm]{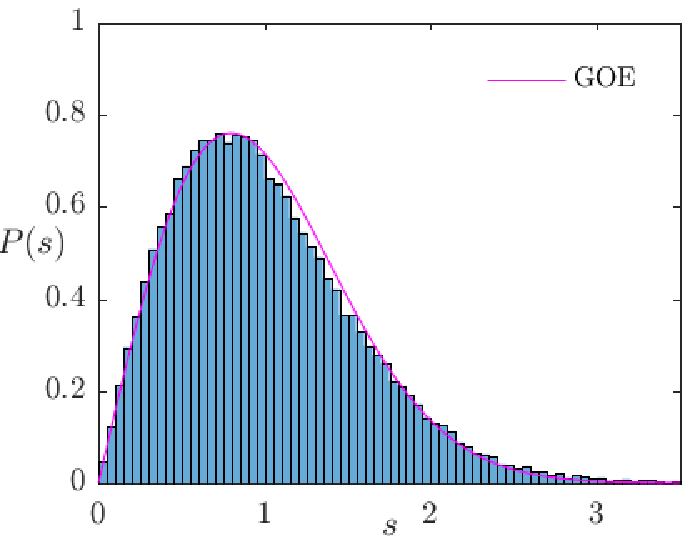}
  }
  \subfloat[][]{
  \centering
  \includegraphics[width=6.5cm]{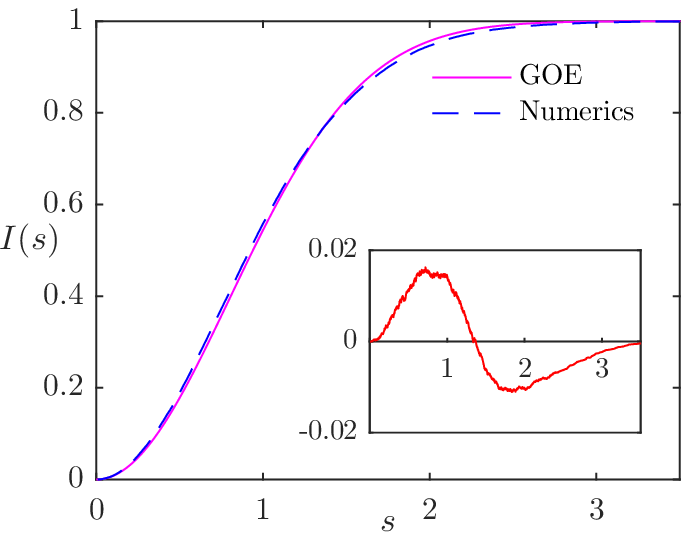}
  }
\caption[]{(a) A histogram of the nearest-neighbor spacing distribution for $50,036$ eigenvalues of the circulant graph $C_{49}(\vec{L};(3,4,9,12,15,19,20))$ with random edge lengths compared to the Wigner surmise for the nearest-neighbor spacing distribution of the eigenvalues of a large random matrix from the GOE. 
 (b) The corresponding integrated nearest-neighbor spacing distribution plotted against the GOE Wigner surmise. The inset shows the difference between the numerical data and the GOE Wigner surmise.}\label{fig:statistics}
\end{figure}

The matrix $M(k)$ in the secular equation of a generic circulant graph (\ref{seculareq}) is an $n \times n$ matrix in contrast to the $2\lv \cE \rv \times 2\lv \cE \rv$ bond scattering matrix that appears in the typical formulation of a secular equation for a quantum graph, see e.g. \cite{BK13}.  This allows one to find the roots efficiently.  Figure \ref{fig:statistics} shows nearest-neighbor spacing statistics for $50,036$ eigenvalues of the circulant graph $C_{49}(\vec{L};(3,4,9,12,15,19,20))$ with the edge lengths uniformly chosen random numbers in $(1,1.5)$.
Figure \ref{fig:statistics}(a) shows a histogram of the nearest-neighbor spacings of the unfolded graph spectrum compared to the distribution of the eigenvalues of large random matrices from the Gaussian Orthogonal Ensemble (GOE).  We see the expected agreement with the random matrix distribution as conjectured for a generic quantum system with time reversal symmetry by Bohigas, Giannoni and Schmit \cite{BGS84} and typically observed in quantum graphs \cite{GS06,KS97,KS99}.
Figure \ref{fig:statistics}(b) plots the integrated nearest neighbor spacing distribution.  An insert shows the difference between the numerical results and the GOE.  The systematic nature of the difference is due to the comparison with the Wigner surmise.

\subsection{Intermediate statistics}

For a circulant graph with symmetric edge lengths the quantum graph spectrum decomposes into subspectra whose eigenfunctions transform according to irreducible representations of $\mathbb{Z}^n$.  In section \ref{sec:intermediate} we showed that the subspectra exhibit intermediate statistics and made predictions for the small and large parameter asymptotics of the two-point correlation function $R_2(x)$, equations \eref{smallxapprox} and \eref{largexapprox}.  The eigenvalues in the subspectra transforming according to the irreducible representation $\cS_j$ correspond to zeros of the function $p_j(k)$ (\ref{pj}).  As $p_j(k)$ is increasing and its asymptotes are known it is straightforward to find the roots, as one root lies between each pair of adjacent asymptotes.

\begin{figure}[htb!]
\centering
  \subfloat[][]{
  \centering
   \includegraphics[width=6.5cm]{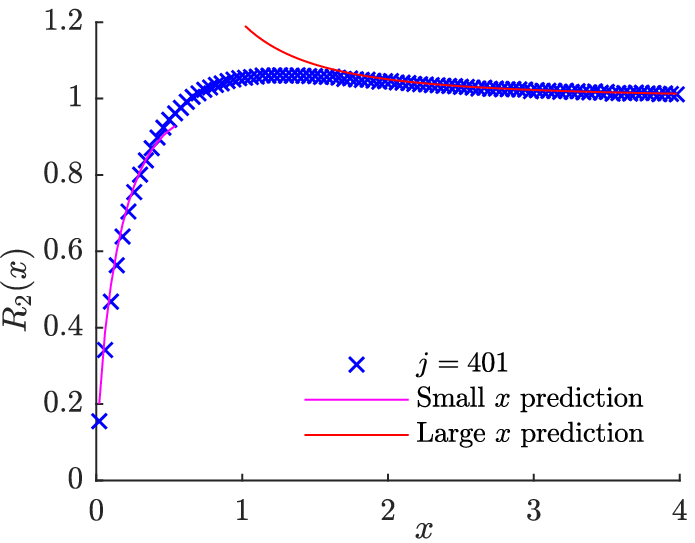}
  }
  \subfloat[][]{
  \centering
  \includegraphics[width=6.5cm]{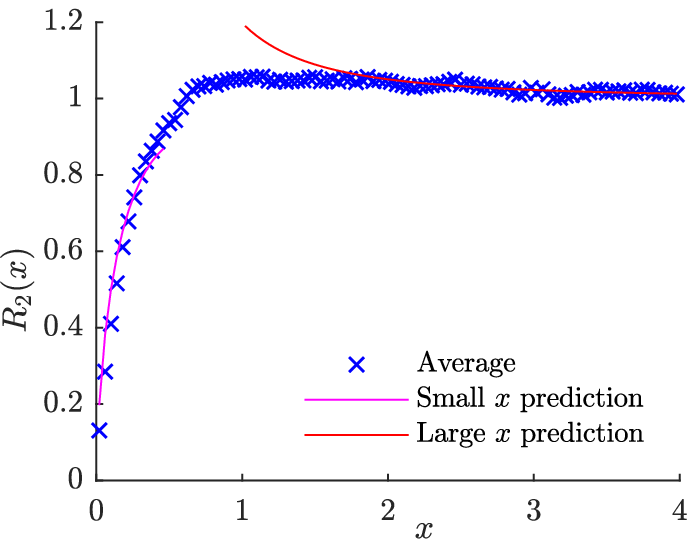}
  }
\caption[]{
(a) The two-point correlation function for a subspectrum of the circulant $C_{7919}(\bell;\vec{a})$ transforming according to the irreducible representation $\cS_{401}$. The vector $\vec{a}$ was randomly generated using a Bernoulli trial on each integer between $1$ and $3959$ with probability $p=1/2$. (b) The two-point correlation function of the subspectra of $C_{7919}(\bell;\vec{a})$ averaged over all non-trivial irreducible representations $\cS_j$ with $j \in \left\{1,\dots,3959\right\}$. }\label{fig:intermediate}
\end{figure}

Predictions for the two-point correlation function were made for randomly generated circulant graphs.  In order to avoid extra Dirichlet eigenvalues identified in lemma \ref{lem:dirichlet} we choose $n$ to be prime and generate the vector $\vec{a}$ via a Bernoulli trial on each integer between $1$ and $(n-1)/2$ with probability $p$.  Figure \ref{fig:intermediate}(a) shows the two-point correlation function for the subspectrum transforming according to the representation $\cS_{401}$ computed from $20,000,089$ eigenvalues of a randomly generated circulant graph $C_{7919}(\bell;\vec{a})$ with $p=1/2$.  We see good agreement with the predictions for both the small and large parameter asymptotics.  The constant $c = 5.5145$ in the small parameter prediction \eref{smallxapprox} was obtained using least squares fitting.  It should be noted that, while the large parameter asymptotics differs from the predictions for star graphs and Dirac rose graphs which also exhibit intermediate statistics, they only differ from the coefficient of the $x^{-4}$ term in the expansion, a slight change in the shape of the large $x$ prediction.    

In the absence of Dirichlet eigenvalues in the subspectra, the predicted behavior of the two-point correlation function is independent of the choice of representation.  
So for comparison we averaged the two-point correlation function calculated from  $5,362$ eigenvalues of each representation for $j=1,\dots 3959$ for a total of $21,228,158$ levels of a random graph $C_{7919}(\bell;\vec{a})$, figure \ref{fig:intermediate}(b).  We see equally good agreement with the asyptotic predictions although calculating the constant in the small parameter prediction using least square fitting, in the averaged case, we find $c = 5.2559$.

\section{Spectral zeta function}\label{sec:zeta}

Many spectral properties are encoded in the spectral zeta function. A spectral zeta function generalizes the Riemann zeta function, replacing the sum over the integers with a sum over the eigenvalues. For a quantum graph with eigenvalues $0 \leq \lambda_0 \leq \lambda_1 \leq \lambda_2 \leq \cdots$ such that 
$\lambda_i = k_i^2$, the associated spectral zeta function is defined as
\begin{equation}\label{spectralzeta}
\zeta(s) = \sideset{}{'} \sum_{i=0}^\infty k_i^{-2s},
\end{equation}
where the prime indicates that any eigenvalues of zero are omitted from the sum. The spectral zeta function for a quantum graph formulated in terms of the vertex conditions was studied in \cite{HK11,HKT12} for the Laplace and Schr{\"o}dinger operators respectively. Here we construct simpler formulas for the zeta function of quantum circulant graphs utilizing the secular equations found  previously.

\subsection[Zeta function with edge symmetry]{The zeta function of circulant graphs with edge symmetry}

For a circulant graph with symmetric edge lengths, recall that the secular equation is the product of the $p_j(k)$ functions defined in \eref{pj}. For a fixed $j$, let $0 < k_{j_0} \leq k_{j_1} \leq k_{j_2} \leq \cdots$ be the positive zeros of $p_j(k)$. For each $j$ we define
\begin{equation}\label{jthzeta}
\zeta_j(s) = \zeta_{j_D}(s) + \sideset{}{'} \sum_{i=0}^\infty k_{j_i}^{-2 s} \ ,
\end{equation}
where $\zeta_{j_D}$ is the sum over the Dirichlet eigenvalues corresponding to eigenfunctions transforming according to the irreducible representation $\cS_j$. Recall, from \sref{sec:dirchlet}, that $\left( m\pi / \ell_h \right)^2$ is in the spectrum whenever $2 j a_h \equiv 0 \Mod{n}$. Then, the spectral zeta function for an edge-symmetric circulant graph is,
\begin{equation}\label{symmetriczeta}
\zeta(s) = \cases{
           \zeta_0(s) + 2\sum_{j=1}^{(n-1)/2} \zeta_j(s) & if $n$ is odd \\
           \zeta_0(s) + \zeta_{n/2}(s) + 2\sum_{j=1}^{(n/2)-1} \zeta_j(s) & if $n$ is even
           } \ .
\end{equation}

\subsubsection{Integral representation of $\zeta_0(s)$}

We employ a contour integral technique introduced in \cite{KM03,KM04} to find an integral representation of the zeta function. To introduce the procedure we start by finding a representation of $\zeta_0(s)$ from $p_0(k)$. Recall that $p_0(k) = 2\sum_{h=1}^d \tan\left(k\ell_h / 2\right)$ and define a function,
\begin{equation}\label{f0z}
f_0(z) = \frac{1}{z} \sum_{h=1}^d \tan\left(\frac{z\ell_h}{2}\right) \ ,
\end{equation}
where $z = k + \rmi t \in \CC$. Then the zeros of $f_0$ are zeros of $p_0$, except for the zero of $p_0$ at the origin, which has been removed.

\begin{figure}[htb!]
\centering
  \subfloat[][]{
    \includegraphics{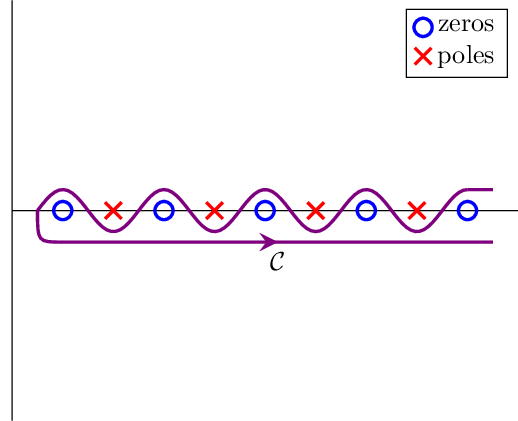}}
  \qquad
  \subfloat[][]{
    \includegraphics{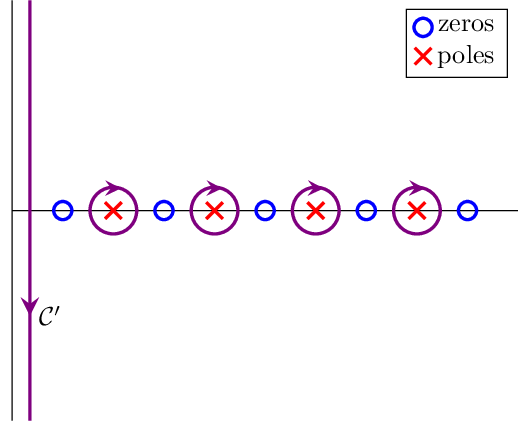}}
\caption[]{(a) The contour $\cC$, which encloses the zeros and avoids the poles of $f_0(z)$. 
(b) The contour $\cC'$, which lies along the imaginary axis and loops around the poles of $f_0(z)$.}\label{fig:contours}
\end{figure}

We use the argument principle to write $\zeta_0(s)$ as a contour integral. Let $\cC$ be a contour which encloses the zeros of $f_0(z)$ and avoids its poles, as shown in figure \ref{fig:contours}(a). Then, 
\begin{eqnarray}
\zeta_0(s) &= \zeta_{0_D}(s) + \frac{1}{2\pi\rmi} \int_\cC z^{-2s} \frac{f_0'(z)}{f_0(z)} \dif z \\
           &= \sum_{m=1}^\infty \sum_{h=1}^d \left(\frac{2m\pi}{\ell_h}\right)^{-2s} + \frac{1}{2\pi\rmi} \int_\cC z^{-2s} \derive{z} \log f_0(z) \dif z \ . \label{contourint}
\end{eqnarray}
We can deform the contour  $\cC$ to a contour $\cC'$, shown in  figure \ref{fig:contours}(b), which lies along the imaginary axis with a loop around each pole. Then it is natural to write
\begin{equation}
\zeta_0(s) = \zeta_{0_P}(s) + \zeta_{0_I}(s) + \zeta_{0_D}(s) \ ,
\end{equation}
where $\zeta_{0_I}(s)$ is the contribution to \eref{contourint} from the integral along the imaginary axis and $\zeta_{0_P}(s)$ is the contribution from the poles  of $f_0(z)$. Applying the residue theorem,
\begin{equation}\label{zeta0Poles}
\zeta_{0_P}(s) = \sum_{m=0}^\infty \left(2m+1\right)^{-2s} \sum_{h=1}^d \left(\frac{\pi}{\ell_h}\right)^{-2s} \ .
\end{equation}
Combining this with  $\zeta_{0_D}$,
\begin{equation}\label{zeta0DP}
\zeta_{0_P}(s) + \zeta_{0_D}(s) = \zeta_R \left(2s\right) \sum_{h=1}^d \left(\frac{\pi}{\ell_h}\right)^{-2s} \ ,
\end{equation}
where $\zeta_R(s)$ is the Riemann zeta function.

Now, consider the integral along the imaginary axis,
\begin{eqnarray}
\zeta_{0_I}(s) &= \frac{1}{2\pi\rmi} \int_{\infty}^{-\infty} (\rmi t)^{-2s} \derive{t} \log f_0(\rmi t) \dif t \\
               &= \frac{\sin\pi s}{\pi} \int_{0}^{\infty} t^{-2s} \derive{t} \log \left(\frac{\hat{f}_0(t)}{t}\right) \dif t \ , \label{zeta0Im}
\end{eqnarray}
where $\hat{f}_0(t) = \sum_{h=1}^d \tanh\left(t\ell_h / 2\right) $. Looking at the asymptotic behavior of $\hat{f}_0(t) / t$, 
\begin{equation}
\frac{\hat{f}_0(t)}{t} \sim 
  \cases{
  \sum_{h=1}^d \left(\frac{\ell_h}{2} - \frac{\ell_h^3 t^2}{24}\right) +\Or(t^4), & as $t \to 0$ \\
  \sum_{h=1}^d \left(\frac{1}{t} + \Or\left(\frac{\rme^{-t\ell_h}}{t}\right)\right), & as $t \to \infty$ \\
  } \ .
\end{equation}
Therefore $\derive{t} \log\hat{f}_0(t) / t$ is proportional to $t$ near zero, and $\derive{t} \log\hat{f}_0(t) / t$ is asymptotic to $1/t$ as $t\to\infty$. Consequently, the integral representation \eref{zeta0Im} holds in the strip $0 < \Re(s) < 1$.

To obtain an analytic continuation of $\zeta_{0_I}(s)$ valid for $\Re(s) \leq 0$, we can split the integral in \eref{zeta0Im} at $t = 1$ and develop the integral over $(1,\infty)$, so
\begin{equation}\label{analyticcontinuation0}
\fl
\zeta_{0_I}(s) = \frac{\sin\pi s}{\pi} \left[\int_0^1 t^{-2s} \derive{t} \log\frac{\hat{f}_0(t)}{t} \dif t + \int_1^\infty t^{-2s} \derive{t} \log\hat{f}_0(t) \dif t - \frac{1}{2s}\right] \ ,
\end{equation}
which converges for $\Re(s) < 1$.

\subsubsection{Integral representation of $\zeta_j(s)$}

For other values of $j$ we follow the same procedure. Let
\begin{eqnarray}
f_j(z) = z\sum_{h=1}^d \left[\cos\left(\frac{2\pi j a_h}{n}\right) \csc(z\ell_h) - \cot(z\ell_h)\right] \label{fjz} \\
\hat{f}_j(t) = \sum_{h=1}^d \left[\coth\left(t\ell_h\right) - \cos\left(\frac{2\pi j a_h}{n}\right) \csch\left(t\ell_h\right)\right] \ , \label{fjhat}
\end{eqnarray}
for each $1 \leq j \leq (n-1) / 2$. Observe that the zeros of $f_j(z)$  again correspond to the zeros of $p_j(k)$, with now a pole removed at the origin. Note that in this case, $f_j(\rmi t) = -t\hat{f}_j(t)$. We represent $\zeta_j$ with a contour integral, 
\begin{equation}
\zeta_{j}(s) = \zeta_{j_D}(s) + \frac{1}{2\pi\rmi} \int_\cC z^{-2s} \derive{z} \log f_j(z) \dif z \ ,
\end{equation}
and again use the contour transformation from $\cC$ to $\cC'$ (See figure \ref{fig:contours}(b)). We may now write
\begin{equation}
\zeta_j(s) = \zeta_{j_P}(s) + \zeta_{j_I}(s) + \zeta_{j_D}(s) \ .
\end{equation}
The integral along the imaginary axis in this case becomes
\begin{equation}\label{zetajIm}
\zeta_{j_I}(s) = \frac{\sin\pi s}{\pi} \int_{0}^{\infty} t^{-2s} \derive{t} \log \left(t\hat{f}_j(t)\right) \dif t \ ,
\end{equation}
which is defined for $0 < \Re(s) < 1$. Again splitting the integral at $t = 1$,
\begin{equation}\label{analyticcontinuationj}
\fl
\zeta_{j_I}(s) = \frac{\sin\pi s}{\pi} \left[ \int_0^1 t^{-2s} \derive{t} \log t\hat{f}_j(t) \dif t + \int_1^\infty t^{-2s} \derive{t} \log\hat{f}_j(t) \dif t + \frac{1}{2s} \right] \ ,
\end{equation}
which is an analytic continuation valid for $\Re(s) < 1$.

To find $\zeta_{j_P}(s)$, we must determine the poles of $f_j(z)$. In general, these occur for $z = m\pi / \ell_h$, since these are roots of $\sin(z\ell_h) = 0$. However, whenever $2 j a_h \equiv 0 \Mod{n}$ a set of poles are removed, but each pole that is removed is replaced by a Dirichlet eigenvalue. For a fixed $j$, define the set
\begin{equation}\label{Hj}
H_j := \left\{1 \leq h \leq d : 2 j a_h \equiv 0 \Mod{n}\right\}
\end{equation}
the set of values of $h$ for which there are Dirichlet eigenvalues rather than poles. 

If $n$ is odd, $n/2$ is not an integer, and for each $h \in H_j$, $j a_h \equiv 0 \Mod{n}$, so we only gain Dirichlet eigenvalues corresponding to even integers. Then,
\begin{equation}\label{zetajOddPoles}
\fl
\zeta_{j_P}(s) = \sum_{m=1}^\infty (2m-1)^{-2s} \sum_{h=1}^d \left(\frac{\pi}{\ell_h}\right)^{-2s} + \sum_{m=1}^\infty (2m)^{-2s} \sum_{\substack{h=1 \cr h \notin H_j}}^d \left(\frac{\pi}{\ell_h}\right)^{-2s} \ ,
\end{equation}
and
\begin{equation}\label{zetajOddDir}
\zeta_{j_D}(s) = \sum_{m=1}^\infty (2m)^{-2s} \sum_{\substack{h=1 \cr h \in H_j}}^d \left(\frac{\pi}{\ell_h}\right)^{-2s} \ .
\end{equation}
Combining \eref{zetajOddPoles} and \eref{zetajOddDir},
\begin{equation}\label{pleasantresult}
\zeta_{j_P}(s) + \zeta_{j_D}(s) = \zeta_R\left(2s\right) \sum_{h=1}^d \left(\frac{\pi}{\ell_h}\right)^{-2s} \ .
\end{equation}

If $n$ is even, it is possible that $j a_h \equiv n/2 \Mod{n}$ or $j a_h \equiv 0 \Mod{n}$. Note that if $j a_h \equiv 0 \Mod{n}$, then $\zeta_{j_P}$ and $\zeta_{j_D}$ are again defined by \eref{zetajOddPoles} and \eref{zetajOddDir}. However, when $j a_h \equiv n/2 \Mod{n}$, we gain Dirichlet eigenvalues corresponding to odd integers. Then,
\begin{equation}\label{zetajEvenPoles}
\fl
\zeta_{j_P}(s) = \sum_{m=1}^\infty (2m)^{-2s} \sum_{h=1}^d \left(\frac{\pi}{\ell_h}\right)^{-2s} + \sum_{m=1}^\infty (2m-1)^{-2s} \sum_{\substack{h=1 \cr h \notin H_j}}^d \left(\frac{\pi}{\ell_h}\right)^{-2s}
\end{equation}
and
\begin{equation}\label{zetajEvenDir}
\zeta_{j_D}(s) = \sum_{m=1}^\infty (2m-1)^{-2s} \sum_{\substack{h=1 \cr h \in H_j}}^d \left(\frac{\pi}{\ell_h}\right)^{-2s} \ .
\end{equation}
Combining \eref{zetajEvenPoles} and \eref{zetajEvenDir} we again obtain \eref{pleasantresult}.

Finally, when $n$ is even, we must consider zeros of the function $p_{n/2}(k)$, which is defined slightly differently from the other $p_j(k)$. Let,
\begin{equation}\label{fn2z}
f_{n/2}(z) = z\left(\sum_{\substack{h=1 \cr a_h\textrm{ is even}}}^d \tan\left(\frac{z\ell_h}{2}\right) - \sum_{\substack{h=1 \cr a_h\textrm{ is odd}}}^d \cot\left(\frac{z \ell_h}{2}\right)\right)
\end{equation}
and
\begin{equation}\label{fn2hat}
\hat{f}_{n/2}(t) = \sum_{\substack{h=1 \cr a_h\textrm{ is even}}}^d \tanh\left(\frac{t\ell_h}{2}\right) + \sum_{\substack{h=1 \cr a_h\textrm{ is odd}}}^d \coth\left(\frac{t\ell_h}{2}\right) \ .
\end{equation}
This case is developed in the same way as $\zeta_0$ in the previous section, ultimately yielding
\begin{equation}
\fl
\zeta_{n/2_I}(s) = \frac{\sin\pi s}{\pi} \left[\int_0^1 t^{-2s} \derive{t} \log t\hat{f}_{n/2}(t) \dif t + \int_1^\infty t^{-2s} \derive{t} \log\hat{f}_{n/2}(t) \dif t + \frac{1}{2s}\right],
\end{equation}
where $\Re(s) < 1$. It can also be shown that the sum of the contributions from the poles and the Dirichlet eigenvalues for $j = n/2$ is given by \eref{pleasantresult}. 

\subsubsection{Full spectral zeta function}

Summing the functions for each $j$ according to \eref{spectralzeta}, we obtain the spectral zeta function of a quantum circulant graph with edge length symmetry.

\begin{thm}\label{thm:symmetriczeta}
  For the Laplace operator on a quantum circulant graph $C_n(\bell,(a_1,\dots,a_d))$ with symmetric edge lengths and standard vertex conditions, the spectral zeta function for $\Re(s) < 1$ is
  \begin{eqnarray}\label{zetafuncodd}
 \fl \zeta(s) = n\zeta_R(2s) \sum_{h=1}^d \left(\frac{\pi}{\ell_h}\right)^{-2s} \cr 
 \fl \qquad+ \frac{\sin\pi s}{\pi}\left[\int_0^1 t^{-2s} \derive{t} \log\frac{\hat{f_0}(t)}{t} \dif t + \int_1^\infty t^{-2s} \derive{t} \log\hat{f_0}(t) \dif t - \frac{1}{2s}\right] \\
\fl  \qquad+ \frac{2\sin\pi s}{\pi} \sum_{j=1}^{(n-1)/2} \left[\int_0^1 t^{-2s} \derive{t} \log t\hat{f_j}(t) \dif t + \int_1^\infty t^{-2s} \derive{t} \log\hat{f_j}(t) \dif t + \frac{1}{2s}\right] \nonumber
  \end{eqnarray}
  when $n$ is odd and 
  \begin{eqnarray}\label{zetafunceven}
  \fl \zeta(s) = n\zeta_R(2s) \sum_{h=1}^d \left(\frac{\pi}{\ell_h}\right)^{-2s} 
+ \frac{\sin\pi s}{\pi}\left[\int_0^1 t^{-2s} \derive{t} \log\frac{\hat{f_0}(t)}{t} \dif t 
\right. \nonumber \\
\fl  +\left.  \int_1^\infty t^{-2s} \derive{t} \log\hat{f_0}(t) \dif t+ \int_0^1 t^{-2s} \derive{t} \log t\hat{f}_{n/2}(t) \dif t 
 + \int_1^\infty t^{-2s} \derive{t} \log\hat{f}_{n/2}(t) \dif t \right] \\
\fl \qquad  + \frac{2\sin\pi s}{\pi} \sum_{j=1}^{(n/2)-1} \left[\int_0^1 t^{-2s} \derive{t} \log t\hat{f_j}(t) \dif t + \int_1^\infty t^{-2s} \derive{t} \log\hat{f_j}(t) \dif t + \frac{1}{2s}\right] \nonumber
  \end{eqnarray}
  when $n$ is even.
\end{thm}

\subsection[Zeta functions of generic circulant graphs]{Spectral zeta functions of generic circulant graphs}

A similar method can be used to obtain the spectral zeta function of a quantum circulant graphs with incommensurate edge lengths. Recall that the secular equation for such a graph is, 
\begin{equation}
\det M(k) = 0 \ ,
\end{equation}
where the $n \times n$ matrix $M(k)$ is defined by
\begin{equation}
\left[M(k)\right]_{ij} =
  \cases{
  -\sum_{v \sim i} \cot k L_{i,v} & $i = j$ \\
  \csc k L_{i,j} & $i \sim j$ \\
  0 & otherwise \\
  }
\end{equation}
see \eref{Mdiag} and \eref{Moffdiag}. 
 
We again define a complex-valued function whose zeros match those of $\det M(k)$ and use the argument principle to express the zeta function as a contour integral. Let
\begin{equation}\label{fgeneralzeta}
f(z) = z^{n-2} \det M(z),
\end{equation}
where $z = k + \rmi t \in \CC$. The zeros of $f$ correspond to the roots of the secular equation, and multiplication by $z^{n-2}$ removes the zero at the origin. The zeta function is expressed as the contour integral,
\begin{equation}
\zeta(s) = \frac{1}{2\pi\rmi} \int_\cC z^{-2s} \derive{z} \log f(z) \dif z \ ,
\end{equation}
where $\cC$ encloses the positive zeros of $f$ and avoids poles. Deforming the contour $\cC$ to $\cC'$ and the zeta function can be written as the sum of residues at the poles of $f$ and an integral along the imaginary axis,
\begin{equation}
\zeta(s) =  \zeta_P(s) + \zeta_I(s) \ .
\end{equation}

The poles of $f$ occur when $z = m\pi / L_{i,j}$, and so
\begin{equation}
\zeta_P(s) = \zeta_R(2s) \sum_{(i,j)\in\cE} \left(\frac{\pi}{L_{i,j}}\right)^{-2s} \ .
\end{equation}
If we define a function $\hat{f}(t) = \det \hat{M}(t)$ where $\hat{M}(t)$ is 
\begin{equation}\label{Mhat}
\left[\hat{M}(t)\right]_{ij} =
  \cases{
  -\sum_{v\sim i} \coth t L_{i,v} & $i = j$ \\
  \csch t L_{i,j} & $i \sim j$ \\
  0 & otherwise
  }
\end{equation}
then $t^{n-2} \hat{f}(t) = f(\rmi t)$. So the integral along the imaginary axis can be written, 
\begin{equation}\label{iaxisgeneralzeta}
\zeta_{I}(s) = \frac{\sin\pi s}{\pi} \int_{0}^{\infty} t^{-2s} \derive{t} \log t^{n-2} \hat{f}(t) \dif t \ .
\end{equation}
Note that $t^n \hat{f}(t) = \det[t\hat{M}(t)] \sim c t^2 + \Or(t^4)$ as $t$ approaches zero, since the expansions of $\coth(t)$ and $\csch(t)$ contain only odd powers and $\det[t\hat{M}(t)]$ tends to zero.  Without loss of generality we can assume $c \neq 0$ as if it was zero for some choice of edge lengths it can be made non zero by an arbitrarily small perturbation. Hence, the integral in \eref{iaxisgeneralzeta} converges for $0 < \Re(s) < 1$. 

To obtain an analytic continuation valid for all $\Re(s) < 1$ we can again split the integral at $t = 1$ and develop the integral over $(1,\infty)$,
\begin{equation}
\fl
\zeta_{I}(s) = \frac{\sin\pi s}{\pi} \left[\int_0^1 t^{-2s} \derive{t} \log t^{n-2}\hat{f}(t) \dif t + \int_1^\infty t^{-2s} \derive{t} \log\hat{f}(t) \dif t + \frac{(n-2)}{2s} \right] \ .
\end{equation}
From this, we obtain the main result of this section.

\begin{thm}\label{thm:generalzeta}
  The spectral zeta function for the Laplace operator on a quantum circulant graph $C_n(\vec{L},(a_1,\dots,a_d))$ with standard vertex conditions, for $\Re(s) < 1$ is
  \begin{eqnarray}\label{zetafuncgen}
  \zeta(s) &= \zeta_R(2s) \sum_{(i,j)\in\cE} \left(\frac{\pi}{L_{i,j}}\right)^{-2s} + \frac{\sin\pi s}{\pi} \left[\int_0^1 t^{-2s} \derive{t} \log t^{n-2}\hat{f}(t) \dif t \right. \nonumber \\
& \left. + \int_1^\infty t^{-2s} \derive{t} \log\hat{f}(t) \dif t + \frac{(n-2)}{2s}\right] \ .
  \end{eqnarray}
\end{thm}

\section{Spectral determinant}\label{sec:determinant}

The spectral determinant of a linear operator is formally the product of its eigenvalues. For the Laplace operator $\Delta$, we can use the zeta function to define a regularized spectral determinant,
\begin{equation}
\det \Delta = \rme^{-\zeta'(0)} \ .
\end{equation}

\subsection[Spectral determinant with edge symmetry]{Spectral determinant of circulant graphs with edge symmetry}

From the zeta function for a quantum circulant graph with edge symmetry, theorem \ref{thm:symmetriczeta}, one can compute the spectral determinant. Taking the derivative of $\zeta(s)$ for odd $n$,
\begin{eqnarray}
\eqalign{
\fl
\zeta'(0) = -d n \log 2\pi + n\sum_{h=1}^d \log\frac{\pi}{\ell_h} - \left.\log\left(\frac{\hat{f}_0(t)}{t} \prod_{j=1}^{(n-1)/2} \lv t\hat{f}_j(t) \rv ^2\right) \rv_{t=0} \cr
 + \left. \log \left(\hat{f}_0(t) \prod_{j=1}^{(n-1)/2} \lv \hat{f}_j(t) \rv^2 \right) \rv_{t\to\infty}
 }  \\
\fl
\qquad = -\log\left[\frac{2^{E-1}\cL}{n d^n} \prod_{h=1}^d \ell_h^n \prod_{j=1}^{(n-1)/2} \left(\sum_{h=1}^d \frac{1}{\ell_h}\left(1 - \cos\frac{2\pi j a_h}{n}\right)\right)^2\right] \ ,
\end{eqnarray}
where $\cL = n\sum_{h=1}^d \ell_h$ is the total length of the graph. The case where $n$ is even follows similarly, giving us the following result.

\begin{thm}
  The spectral determinant of the Laplace operator of a quantum circulant graph $C_n(\bell;\vec{a})$ with symmetric edge lengths and standard vertex conditions is
  \begin{equation}
   \det \Delta = 
    \frac{2^{E-1} \cL}{n d^n} \prod_{h=1}^d \ell_h^n \prod_{j=1}^{(n-1)/2} \left(\sum_{h=1}^d \frac{1}{\ell_h} \left(1-\cos\frac{2\pi j a_h}{n}\right)\right)^2 
    \end{equation}
     when $n$ is odd and
    \begin{equation}
    \fl \det \Delta = 
    \frac{2^{E-1} \cL}{n d^n} \prod_{h=1}^d \ell_h^n \prod_{j=1}^{(n/2)-1} \left(\sum_{h=1}^d \frac{1}{\ell_h} \left(1-\cos\frac{2\pi j a_h}{n}\right)\right)^2 \left(\sum_{\substack{h=1 \cr a_h\textrm{ is odd}}}^d \frac{2}{\ell_h}\right) 
    \end{equation}
    when $n$ is even.  $d$ is the number of elements in $\vec{a}$, $E = nd$ is the number of edges, and $\cL = n\sum_{h=1}^d \ell_h$ is the total length of the graph.
\end{thm}

\subsection{Spectral determinant of generic circulant graphs}

Similarly we can use theorem \ref{thm:generalzeta} to compute the spectral determinant of a generic quantum circulant graph.

\begin{thm}
  The spectral determinant of the Laplace operator of a quantum circulant graph $C_n(\vec{L};\vec{a})$ with standard vertex conditions is
  \begin{equation}\label{spectraldetstandard}
  \det \Delta = c \left(-\frac{2^{d-1}}{d}\right)^n \prod_{(i,j)\in\cE} L_{i,j}
  \end{equation}
  where $d$ is the number of elements in $\vec{a}$ and $c$ is the first nonzero coefficient in the expansion of $\det[t\hat{M}(t)]$ near zero.
  \begin{proof}
    Note that
    \begin{equation}
\fl    \zeta'(0) = -nd \log 2\pi + \sum_{(i,j)\in\cE} \log\frac{\pi}{L_{i,j}} - \left. \log\left(t^{n-2}\hat{f}(t)\right) \rv_{t=0} + \left. \log\hat{f}(t) \rv_{t\to\infty}
    \end{equation}
    where $\hat{f}(t) = \det\hat{M}(t)$ as defined in \eref{Mhat}. As $t\to\infty$, $\hat{M}(t) \sim -2d \id_n$, and so $\hat{f}(t) \sim (-2d)^n$.
    
    To evaluate $t^{n-2} \hat{f}(t)$ as $t\to 0$, recall that $t^n \det\hat{M}(t) \sim c t^2 + \Or(t^4)$, since the constant term in the expansion must be zero. The first nonzero coefficient in the expansion is $c$. Note that the expansion is straightforward to compute for any given graph.
  \end{proof}
\end{thm}

\section{Vacuum energy}\label{sec:vacuum}

Another application of the spectral zeta function is to define a regularized vacuum (Casimir) energy associated with the graph \cite{HK11, HKT12}. Graph vacuum energy was also studied in \cite{BHW09,FKW07}. The vacuum energy is formally a sum of the square roots of the eigenvalues of the Laplacian,
\begin{equation*}
\frac{1}{2} \sum_{j=0}^\infty \sqrt{\lambda_j}  \ .
\end{equation*}
The zeta regularized vacuum energy is then defined as,
\begin{equation}
E_c = \frac{1}{2} \zeta\left(-\frac{1}{2}\right) \ .
\end{equation}

\subsection[Vacuum energy with edge symmetry]{Vacuum energy of circulant graphs with edge symmetry}

We use the zeta function in theorem \ref{thm:symmetriczeta} to compute the vacuum energy of a symmetric circulant graph.

\begin{thm}
  The vacuum energy of the Laplace operator of a quantum circulant graph $C_n(\bell;\vec{a})$ with symmetric edge lengths and standard vertex conditions is,
  \begin{eqnarray}\label{casimirforcesym}
  \fl
  E_c =
        \frac{n-2}{2\pi} - \frac{n\pi}{24} \sum_{h=1}^d (\ell_h)^{-1} - \frac{1}{2\pi} \left[\int_0^1 t \derive{t} \log\left(t^{n-2}\hat{f_0}(t) \prod_{j=1}^{(n-1)/2} \left[\hat{f_j}(t)\right]^2\right) \dif t \right. \nonumber \\
+ \left. \int_1^\infty t \derive{t} \log\left(\hat{f_0}(t) \prod_{j=1}^{(n-1)/2} \left[\hat{f_j}(t)\right]^2\right) \dif t \right] 
\end{eqnarray}
when $n$ is odd and
  \begin{eqnarray}\label{casimirforcesym}
  \fl E_c =      
        \frac{n-2}{2\pi} - \frac{n\pi}{24} \sum_{h=1}^d (\ell_h)^{-1} - \frac{1}{2\pi} \left[\int_0^1 t \derive{t} \log\left(t^{n-2}\hat{f_0}(t)\hat{f}_{n/2}(t) \prod_{j=1}^{(n-1)/2} \left[\hat{f_j}(t)\right]^2\right) \dif t \right. \nonumber \\
 + \left. \int_1^\infty t \derive{t} \log\left(\hat{f_0}(t)\hat{f}_{n/2}(t) \prod_{j=1}^{(n-1)/2} \left[\hat{f_j}(t)\right]^2\right) \dif t \right]
  \end{eqnarray}
when $n$ is even.
\end{thm}

\subsection{Vacuum energy of generic circulant graphs}

Similarly the zeta function of a generic circulant graph theorem \ref{thm:generalzeta} provides an integral formulation for the vacuum energy of a generic circulant graph.

\begin{thm}\label{thm:casimirforcestandard}
  The vacuum energy of the Laplace operator of a quantum circulant graph $C_n(\vec{L};\vec{a})$ with standard vertex conditions is
  \begin{equation}
  \fl
  E_c = \frac{n-2}{2\pi} - \frac{\pi}{24} \sum_{(i,j)\in\cE} \left(L_{i,j}\right)^{-1} - \frac{1}{2\pi} \left[\int_0^1 t \derive{t} \log t^{n-2}\hat{f}(t) \dif t + \int_1^\infty t \derive{t} \log\hat{f}(t) \dif t \right]
  \end{equation}
\end{thm}

Comparing this result with the one obtained in \eref{casimirforcesym}, we can immediately see that \eref{casimirforcesym} is a special case of theorem \ref{thm:casimirforcestandard}, when $\hat{f}(t) = \det\hat{M}(t) = \prod_{j=0}^n \hat{f}_j(t)$.

\section{Conclusions}\label{sec:conclusions}

We introduced a class of quantum graphs, quantum circulant graphs, that generalize the widely studied class of quantum star graphs.  The spectrum of a quantum circulant graph is encoded in a secular equation, which takes two forms, depending on whether  the metric respects the symmetry of the circulant graph.  Both forms of the secular equation preserve analogies with the prototypical secular equation of the Neumann star graph.
For a circulant graph with a symmetric metric, the secular equation can also be obtained using a quotient graph technique that was employed in a number of recent articles to identify graphs with unusual spectral properties \cite{BPB09, JMS14, PB10}.

While the spectral statistics of a large quantum circulant graph are seen to agree with random matrix statistics, according to the Bohigas-Giannoni-Schmidt conjecture \cite{BGS99}, the spectral statistics where the edge lengths respect the cyclic symmetry of the graph fall in a class of intermediate statistics.  The particular form of intermediate statistics differs both from that investigated for the star graph and  \v Seba billiards \cite{BK99, BGS99, BGS01sing}, and from the intermediate statistics of the Dirac rose graph, with which the analysis has a close analogy \cite{HW12}.  

To demonstrate the capacity of the secular equations, we developed spectral zeta functions for circulant graphs with and without edge symmetry.  The spectral zeta functions provide new results for the spectral determinant and vacuum energy of the families of circulant graphs.
In addition, many results have been obtained for star graphs that have not been reproduced on generic graphs by exploiting the simple form of the secular equation, see e.g. \cite{BBK01,BK99,BKW04}.  The secular equations for both formulations of circulant graphs maintain important properties of the star graph formulation.  This suggests that further techniques developed to prove results on star graphs may be extended to obtain results for the larger and more generic families of circulant graphs.


\ack
The authors would like to thank Gregory Berkolaiko, Klaus Kirsten and Jens Markof for helpful comments. JH would like to thank the University of Warwick for their hospitality during his sabbatical where some of the work was carried out. JH was supported by the Baylor University research leave program. This work was partially supported by a grant from the Simons Foundation (354583 to Jonathan Harrison).


\end{document}